\documentclass[preprint,sort&compress,times,fleqn]{elsarticle}

\usepackage{amssymb}
\usepackage{amsmath}
\usepackage{graphicx}
\usepackage{multirow}
\usepackage{subfigure}
\usepackage{float}
\usepackage{graphics}
\usepackage{slashed}
\usepackage{booktabs}
\usepackage{color}
\usepackage[normalem]{ulem}
\usepackage{bm,psfrag}
\usepackage{float}
\usepackage[figuresright]{rotating}

\usepackage{bm}
\usepackage{array}
\usepackage{hyperref}
\usepackage{slashed}
\usepackage{braket}
\usepackage{bigstrut}
\usepackage[nooneline,footnotesize,center]{caption2}

\journal{Nuclear Physics A, published in Nucl.Phys. A961 (2017) 154-168 }
\begin{document}

\begin{frontmatter}

\title { Effect of sea quarks on  the single-spin asymmetries $A^{W^{\pm}}_{L}$ in polarized pp collisions at RHIC}

\author[PKU]{Fang Tian}

\author[PKU]{Chang Gong}

\author[PKU,CIC,CHEP]{Bo-Qiang Ma\corref{cor1}}
\ead{mabq@pku.edu.cn}
\cortext[cor1]{Corresponding author at:School of Physics,Peking University,Beijing 100871,China.}

\address[PKU]{School of Physics and State Key Laboratory of Nuclear Physics and
Technology, Peking University, Beijing 100871,
China}
\address[CIC]{Collaborative Innovation Center of Quantum Matter, Beijing, China}
\address[CHEP]{Center for High Energy Physics, Peking University, Beijing 100871, China}

\begin{abstract}
We calculate the single-spin asymmetries $A^{W^{\pm}}_{L}$ of $W^{\pm}$ bosons produced in polarized pp collisions with the valence part of the up and down quark helicity distributions modeled by the light-cone quark-spectator-diquark model while the sea part helicity distributions of the up and down quarks treated as parametrization. Comparing our results with those from experimental data at RHIC, we find that the helicity distributions of sea quarks play an important role in the determination of the shapes of $A^{W^{\pm}}_{L}$. It is shown that $A^{W^{-}}_{L}$ is sensitive to $\Delta \bar u$, while $A^{W^{+}}_{L}$ to $\Delta \bar d$ intuitively. The experimental data of the polarized structure functions and the sum of helicities are also important to constrain the sizes of quark helicity distributions both for the sea part and the valence part of the nucleon.
\end{abstract}

\end{frontmatter}

\section{Introduction}\label{sec:intro}

The spin of the nucleon is an important research frontier in high energy physics. In the late 1980's, the EMC Collaboration~\cite{Ashman:1987hv,Ashman:1989ig} found that the spin of quarks contributes only a small part of the total proton spin through their polarized deep inelastic scattering (DIS) experiments of charged lepton (muon in their case) beams on polarized proton targets. This observation is different from the naive quark model where the spin of the nucleon comes from the sum of its composite quarks. This inspired the so-called ``spin crisis" of how the spin of the nucleon is distributed among its composite partons. Thus the spin physics, especially the spin-dependent parton distribution functions (PDFs) of nucleons and the polarized structure functions of $g^{p}_{1}$ for protons and $g^{n}_{1}$ for neutrons, have received lots of attentions by the SLAC experiments~\cite{Anthony:1993uf,Abe:1994cp} and the NMC experiments~\cite{Ashman:1987hv,Ashman:1989ig,Adeva:1993km,Adams:1994zd}.
The experiments lead to extended deep-inelastic scattering (DIS) data, from which one may obtain an improved result of about 30\% of the proton spin coming from quarks. The new experiments by the COMPASS Collaboration \cite{Alekseev:2007vi,Adolph:2012vj,Adolph:2012ca,Adolph:2015saz} and the HERMES Collaboration \cite{Airapetian:2004zf,Airapetian:2006vy}
can provide precise information to study the spin
structure of nucleons and the quark PDFs inside the nucleon.  The most up-to-date experimental data have been impressively enriched
from various experiments including the semi-inclusive DIS (SIDIS) in fixed target experiments.
The limitation of the DIS data is that only the quark PDF combinations $\Delta q^{+}=\Delta q+\Delta \bar{q}$ are accessible. Though the semi-inclusive deep-inelastic scattering (SIDIS) experiments can serve to separate different contributions from quarks and anti-quarks, there are also limitation due to some assumptions~\cite{Adeva:1997qz,Adeva:2004dh}. Therefore both DIS and SIDIS data are insensitive to the polarized anti-quark distribution functions.

The Drell-Yan process, especially the mid-state of $W$ boson production in polarized proton-proton collisions can serve as a direct and precise tool to extract the polarized anti-quark distribution functions, due to that $W$ boson is produced through $V$-$A$ interaction~\cite{Bourrely:1980mr,Craigie:1984tk,Bourrely:1993dd,Bourrely:1994sc,Bourrely:1995fw}. The quarks have exact helicity when they couple with $W$ boson in the weak interaction.
What is more, the $W$ boson decay has a clean final state without fragmentation process.  When we study the single-spin asymmetry, it can allow unique and useful measurements of the spin distributions of quarks and antiquarks in the proton~\cite{Nadolsky:2003ga,deFlorian:2010aa,Page:2015woa}.
In recent years, the Relativistic Heavy Ion Collider (RHIC) at Brookhaven National Laboratory~\cite{Bourrely:1980mr,Craigie:1984tk,Bourrely:1993dd,Bourrely:1994sc,Bourrely:1995fw,Adare:2008qb,Adare:2008aa,Adare:2010xa,Aggarwal:2010vc,Adamczyk:2012qj,Adamczyk:2013yvv,Adamczyk:2014ozi,Adamczyk:2014xyw,Adare:2014hsq,Adare:2015gsd} provides a direct probe of helicity distributions with the detection of the longitudinal polarized single-spin asymmetry of $W$ boson by the PHENIX \cite{Adare:2008qb,Adare:2008aa,Adare:2010xa,Adare:2014hsq} and the STAR \cite{Aggarwal:2010vc,Adamczyk:2012qj,Adamczyk:2013yvv,Adamczyk:2014ozi,Adamczyk:2014xyw} experiments at RHIC. In~\cite{Aggarwal:2010vc}, the STAR Collaboration reported the first measurement of the parity violating single-spin asymmetries for midrapidity $W$ decay with $\sqrt{s}=500$ GeV in $\vec{p}p$ collisions. In~\cite{Adare:2010xa}, the PHENIX Collaboration presented data from longitudinal polarized pp collision, where the transverse momentum of final lepton satisfies $p_{T}>30$ GeV mainly from $W$ and $Z$ decays with $\sqrt{s}=500$ GeV. The PHENIX Collaboration also released precise data collected in 2011-2013 with a higher integrated luminosity~\cite{Adare:2015gsd}.

There are also some theoretical studies on $A^{W^{\pm}}_{L}$ based on available extractions or parametrizations of quark helicity distributions~\cite{deFlorian:2008mr,deFlorian:2009vb,Nocera:2014gqa,deFlorian:2014yva}.
The predictions of $A^{W^{\pm}}_{L}$~\cite{Bourrely:1980mr,Craigie:1984tk,Bourrely:1993dd,Bourrely:1994sc,Bourrely:1995fw,Bunce:2000uv,Nadolsky:2003ga,Chen:2005jsa,deFlorian:2010aa} are given, and it is found that the higher order QCD corrections are small~\cite{Gehrmann:1997ez,Kamal:1997fg,vonArx:2011fz,Ringer:2015oaa}. Especially  in Ref.~\cite{Ringer:2015oaa}, the analytic expressions for the spin-dependent asymmetries at next-to-leading order are given and the calculated results with sea quark helicity distributions from different groups are compared with the experimental data~\cite{Adamczyk:2014xyw}. It is found that the data prefer a rather sizable positive valued $\Delta \bar u$.

In this paper, we investigate the contribution of sea quark helicity distributions to $A^{W^{\pm}}_{L}$ with the valence quark helicity distributions modeled by the quark-spectator-diquark model. Constraints due to the polarized structure functions and the sum of the helicity distributions are also considered. Sec.~\ref{sec:ww1} presents the necessary formulas of the quark-diquark model. Sec.~\ref{sec:ww2} presents the extractions of sea quark helicity distributions from the corresponding single spin asymmetries of $W^{\pm}$ bosons. We find that the shape of $A^{W^{-}}_{L}$ is sensitive to $\Delta \bar u$, while $A^{W^{+}}_{L}$ to $\Delta \bar d$ intuitively, so that $\Delta \bar u$ is positive valued while $\Delta \bar d$ is negative valued for better description of experimental data. However, the sizes of $\Delta \bar u$ and $\Delta \bar d$ are strongly constrained by the experimental data of polarized structure functions and the sum of helicities. Numerical results and discussions are presented. A summary is given in the final section.

\section{Light-cone quark-spectator-diquark model}\label{sec:ww1}

As well known, the quark-diquark model~\cite{Feynman:1969ej,Bjorken:1968dy,Bjorken:1969ja,Carlitz:1975bg,Kaur:1977ce,Ma:1986ce,Schaefer:1988xs} is proper to describe the deep inelastic scattering processes as that a single constituent quark in the nucleon is struck by the incident lepton, while the remaining part of the struck nucleon is regarded as an effective spectator with diquark quantum numbers. The light-cone quark-spectator-diquark model (qD model)~\cite{Ma:1996np} is a revised version of the quark-diquark model in the light-cone formalism.

The unpolarized valence quark distributions of flavors $q=u$ and $d$ in this model are:
\begin{eqnarray}
u_v(x)&=&\frac{1}{2}a_S(x)+\frac{1}{6}a_V(x),\\
d_v(x)&=&\frac{1}{3}a_V(x),
\label{eq:q1}
\end{eqnarray}
where the superscript ``$v$" denotes the valence part, $a_{D}(x)$ ($D = S$ for scalar spectator or $V$ for axial vector spectator), denoting the amplitude for quark $q$ to be scattered while the spectator is in the diquark state $D$, is expressed as:
\begin{equation}
a_{D}(x) \propto  \int\left[\rm{d}^2 {\mathbf k}_\perp\right] |\varphi (x,
{\mathbf k}_\perp)|^2 \hspace{0.2cm}(D=S \hspace{0.1cm} \mathrm{or}
\hspace{0.1cm} V),
\label{eq:q2}
\end{equation}
and the normalization satisfies $\int_0^1 {\mathrm d} x a_D(x)=3$ as there are 3 valence quarks to serve as the struck quark in the nucleon.

The relation between the helicity distributions in the light-cone frame and the spin distributions in the rest frame is~\cite{Ma:1991xq,Ma:1992sj}
\begin{equation}
\Delta q (x)
=\int [{\rm d}^2{\bf k}_{\perp}] W_D(x,{\bf k}_{\perp})
[q_{s_z=\frac{1}{2}}
(x,{\bf k}_{\perp})-q_{s_z=-\frac{1}{2}}(x,{\bf k}_{\perp})],
\end{equation}
where
\begin{equation}
W_D(x,{\bf k}_{\perp})=\frac{(k^+ +m)^2-{\bf k}^2_{\perp}}
{(k^+ +m)^2+{\bf k}^2_{\perp}},
\end{equation}
is the Melosh-Wigner rotation factor~\cite{Ma:1996np,Ma:1991xq,Ma:1992sj,Ma:1997gy,Ma:1998ar} from the relativistic effect due to
the quark transversal motions,
$q_{s_z=\frac{1}{2}}(x,{\bf k}_{\perp})$
and $q_{s_z=-\frac{1}{2}}(x,{\bf k}_{\perp})$ are the probabilities
of finding a quark and an antiquark with rest mass $m$
and with spin parallel and anti-parallel to the rest proton
spin, and $k^+=x {\cal M}$ where
${\cal M}^{2}=\frac{m^2_q+{\bf k}^2_{\perp}}{x}
+\frac{m^2_D+{\bf k}^2_{\perp}}{1-x}$.
The Wigner rotation factor $W_D(x,{\bf k}_{\perp})$ ranges
from 0 to 1.

With the Wigner rotation effect, we can obtain the valence quark helicity distributions for the up and down quarks~\cite{Ma:1996np}:
\begin{eqnarray}
\Delta u_v(x)&=&\frac{1}{2}a_S(x)W_S(x)-\frac{1}{18}a_V(x)W_V(x),\\
\Delta d_v(x)&=&-\frac{1}{9}a_V(x)W_V(x).
\label{eq:q3}
\end{eqnarray}

For the light-cone momentum space wave function $\varphi_D (x,k_{\perp})$, we adopt the
Brodsky-Huang-Lepage (BHL) prescription~\cite{Brodsky:1981jv,Huang:1994dy}:
\begin{equation}
\varphi_{D} (x, {\mathbf k}_\perp) = A_D \exp \left\{-\frac{1}{8\beta_D^2}
\left[\frac{m_q^2+{\mathbf k}_\perp ^2}{x} + \frac{m_D^2+{\mathbf
		k}_\perp^2}{1-x}\right]\right\},
\end{equation}
Here, $m_q$ is the mass of quark, and $\beta_D$ is the harmonic oscillator scale parameters which are adjustable. We first adopt $m_q=330$~MeV and $\beta_D=330$~MeV, with those parameters it can reproduce the low energy properties as shown in Ref.~\cite{Ma:1996np}. We take $m_S = 600$~MeV and $m_V =800$~MeV for the scalar and vector diquarks to explain the $N$-$\Delta$ mass difference.

By Eqs.~({\ref{eq:q1}}) and ({\ref{eq:q3}}), the relations between the polarized and unpolarized parton distribution functions are:
\begin{eqnarray}
\Delta u_v(x)&=&[u_v(x)-\frac{1}{2}d_v(x)]W_S(x)-\frac{1}{6}d_v(x)W_V(x),\\
\Delta d_v(x)&=&\textcolor[rgb]{1.00,0.00,0.00}{-}\frac{1}{3}d_v(x)W_V(x).
\label{eq:q4}
\end{eqnarray}

In order to reproduce the experimental data in a reasonable form with relations of valence quark distributions in the theoretical qD model being kept, we may adopt the following parametrization:
\begin{eqnarray}
u_v^{\mathrm{para}}(x)&=&u_v^{\mathrm{CT14LO}}(x),\nonumber \\
d_v^{\mathrm{para}}(x)&=&\frac{d_v^{\mathrm{qD}}(x)}{u_v^{\mathrm{qD}}(x)}\times u_v^{\mathrm{para}}(x),\nonumber \\
\Delta u_v^{\mathrm{para}}(x)&=&[u_v^{\mathrm{para}}(x)-\frac{1}{2}d_v^{\mathrm{para}}(x)]\times W_{S}(x)-\frac{1}{6}d_v^{\mathrm{para}}(x)\times W_{V}(x),\nonumber \\
\Delta d_v^{\mathrm{para}}(x)&=&-\frac{1}{3}d_v^{\mathrm{para}}(x)\times W_{V}(x),
\label{eq:q5}
\end{eqnarray}
where the superscript ``CT14LO" means the direct CTEQ parametrization~\cite{Dulat:2015mca}, and ``qD" means the pure theoretical calculation from the qD model. In this way of parametrization, the unpolarized sea distributions could be included as those of the input parametrization. What is more, the sea part and the valence part of quark distributions are consistent with each other. The parton distribution functions (PDFs) are also reasonably scale dependent as they are mainly based on the parametrization set. Thus we get an adjusted set of quark distributions for both unpolarized and polarized cases based on theoretical considerations.

Using a different unpolarized PDF parametrization set as input can change the valence $u$ quark and sea quark distributions. Due to the dominance of $u$ quarks in proton, there exists little difference between different parametrizations for the valence $u$ quarks. Besides, the unpolarized $d$ valence quark and the polarized $u,d$ valence quarks are obtained through the model calculations as in Eq.~(\ref{eq:q5}), and can keep stable with different unpolarized PDF parametrization sets as inputs. While for the polarized sea parts, we assume simple relations between unpolarized and polarized PDFs as in Eqs.~(\ref{eq:01}) and (\ref{eq:02}). So adopting different unpolarized sea PDF parametrizations as inputs makes small difference of the results in our paper. We adopt CTEQ parametrization~\cite{Dulat:2015mca} as an example for the input parametrization.

\section{Polarized sea quark distributions}\label{sec:ww2}
\subsection{Single-spin asymmetry in $W^{\pm}$ boson production}
For a longitudinally polarized $\vec{p}p\rightarrow W^{\pm}+X$ process, the single-spin asymmetry can be defined as:
\begin{equation}
\begin{aligned}
A^{W^{\pm}}_{L}=\frac{d\sigma^{+}-d\sigma^{-}}{d\sigma^{+}+d\sigma^{-}}=\frac{d\Delta\sigma}{d\sigma},
\end{aligned}
\label{eq:0}
\end{equation}
where the superscripts ``$+/-$" mean the helicity directions of the incoming proton. ``$+$" implies that the direction of spin is along the movement of the proton, and ``$-$" means the opposite. $d\Delta\sigma$ and $d\sigma$ are the polarized and unpolarized hadronic cross sections. According to the factorization~\cite{Collins:1989gx}, the hadronic cross section $d\sigma$ can be expressed by the convolution integrals of the related parton distributions and the perturbative partonic cross section $d\hat{\sigma}$ at the factorization scale $\mu_f$ as:
\begin{equation}
d\sigma=\sum_{a,b}\int dx_adx_bf_a(x_a,\mu_f)f_b(x_b,\mu_f)d\hat{\sigma}(x_aP_A,x_bP_B,\mu_f),
\end{equation}
where $f_{a,b}(x_{a,b})$ means the parton distribution function in the proton. $P_A$ and $P_B$ are the momenta of the initially incoming protons. $x_a$ and $x_b$ are the momentum fractions of parent hadrons carried by the scattering partons. For the polarized situations, the cross section $d\Delta\sigma$ can be obtained using the polarized parton distribution functions $\Delta f_{a,b}(x_{a,b})$ and the corresponding polarized partonic cross section $d\Delta \hat{\sigma}$.

The proton is composed mainly by $u$ and $d$ quarks. So the production of the $W$ boson is dominated by $u$ and $d$ contributions. At leading order~(LO), for $u\bar{d}{\rightarrow W}^{+}$, $A^{W^{+}}_{L}$ can be expressed roughly as~\cite{Bourrely:1993dd,Bourrely:1994sc,Bourrely:1995fw,Bunce:2000uv,Chen:2005jsa}:
\begin{equation}
\begin{aligned}
A^{W^{+}}_{L}=\frac{-\Delta u(x_1)\bar{d}(x_2)+\Delta\bar{d}(x_1)u(x_2)}{u(x_1)\bar{d}(x_2)+\bar{d}(x_1)u(x_2)}.
\end{aligned}
\label{eq:1}
\end{equation}
As for $d\bar{u}\rightarrow W^{-}$, $A^{W^{-}}_{L}$ is:
\begin{equation}
\begin{aligned}
A^{W^{-}}_{L}=\frac{-\Delta d(x_1)\bar{u}(x_2)+\Delta\bar{u}(x_1)d(x_2)}{d(x_1)\bar{u}(x_2)+\bar{u}(x_1)d(x_2)},
\end{aligned}
\label{eq:2}
\end{equation}
where the parton momentum fraction $x_{n}~(n=1,2)$ can be determined by the center-of-mass energy $\sqrt{s}$, the rapidity $y_W$ and the mass $M_W$ of $W$ boson as:
\begin{equation}
\begin{aligned}
x_1&=&\frac{M_W}{\sqrt{s}}{e^{y_W}},\\
x_2&=&\frac{M_W}{\sqrt{s}}{e^{-y_W}}.
\end{aligned}
\label{eq:3}
\end{equation}
Then we are able to connect the measured single-spin asymmetry with the quark or anti-quark helicity distributions in the proton.

For $\vec{p}+p\rightarrow l+X$ at RHIC~\cite{Adamczyk:2014xyw}, the momenta of incoming protons and the produced charged lepton can be denoted by $P_A$, $P_B$, and $p_l$. $\eta_l$ and $\vec{p}_T$ are the rapidity and transverse momentum of the final lepton. According to Eq.~(\ref{eq:0}), the final expression related to the experiment is:
\begin{equation}
A_{L}(\eta_l)=\frac{\int d^2 \vec{p}_T d\Delta\sigma}{\int d^2 \vec{p}_T d\sigma},\\
\label{eq:4}
\end{equation}
and the relatively concrete expression can be obtained in~\cite{Adamczyk:2014xyw}.

Additionally, the spin content of the nucleon can be served as another test of our calculations.
The spin-dependent structure functions $g^{p,n}_{1}(x)$ are of fundamental importance in understanding the quark spin structure of the nucleon. The first moments $\Gamma^{p,n}_{1}=\int_{0}^{1}g^{p,n}_{1}(x )dx$ are related to the net quark helicities in the nucleon.
For the proton and the neutron:
\begin{eqnarray}
 \Gamma^{p}_{1}&=&\int_{a}^{1}dx(\frac{2}{9}(\Delta u(x)+\Delta \bar u(x))+\frac{1}{18}(\Delta d(x)+\Delta \bar d(x))),\\
\Gamma^{n}_{1}&=&\int_{a}^{1}dx(\frac{2}{9}(\Delta d(x)+\Delta \bar d(x))+\frac{1}{18}(\Delta u(x)+\Delta \bar u(x))),
\label{eq:03}
\end{eqnarray}
whereas the corresponding contribution from only valence quarks are	
\begin{eqnarray}
\Gamma^{p}_{1v}&=&\int_{a}^{1}dx(\frac{2}{9}(\Delta u(x)-\Delta \bar u(x))+\frac{1}{18}(\Delta d(x)-\Delta \bar d(x))),\\
\Gamma^{n}_{1v}&=&\int_{a}^{1}dx(\frac{2}{9}(\Delta d(x)-\Delta \bar d(x))+\frac{1}{18}(\Delta u(x)-\Delta \bar u(x))).
\end{eqnarray}		
The sums of helicity distributions for the nucleon are:
 \begin{eqnarray}
 \Delta \Sigma&=&\int_{a}^{1}dx(\Delta u(x)+\Delta \bar u(x)+\Delta d(x)+\Delta \bar d(x)),\\
 \Delta \Sigma_{v}&=&\int_{a}^{1}dx(\Delta u(x)-\Delta \bar u(x)+\Delta d(x)-\Delta \bar d(x)),\\
 \Delta q^{+}&=&\int_{a}^{1}dx(\Delta q(x)+\Delta \bar q(x)),
 \label{eq:04}
 \end{eqnarray}		
 where the subscript ``$v$" means the valence part, and $\Delta \bar q(x)$ and $\Delta q(x)$ are the helicity distributions of anti-quarks and quarks. In our paper, all of the valence part are from model calculations in Eq.~(\ref{eq:q5}), while the unpolarized sea distributions are from CTEQ14  parametrization~\cite{Dulat:2015mca}.

The polarized and the unpolarized sea quark distributions should satisfy a general relation $|\Delta \bar{q}(x)|\le \bar{q}(x)$. We thus propose a simple linear (Linear) relation between the polarized and the unpolarized sea distributions as:
\begin{eqnarray}
\begin{aligned}
\Delta \bar{q}(x)=N_{\bar{q}}\bar{q}(x),~~~ {q=u~\mathrm{or}~d},
\label{eq:01}
\end{aligned}
\end{eqnarray}
where $N_{\bar{q}} \le 1$ are free parameters. According to the Pauli principle~\cite{deFlorian:2009vb}, $N_{\bar{u}}\ge 0$ and $N_{\bar{d}}\le 0$.

By extending from the relation Eq.~(\ref{eq:01}), we write down a $x$-dependent non-linear (NLinear) formula for the sea quark helicity distributions,
\begin{eqnarray}
\begin{aligned}
\Delta \bar{q}(x)=n_{\bar{q}}\frac{\Gamma(a_{\bar{q}}+b_{\bar{q}}+2)}{\Gamma(a_{\bar{q}}+1)\Gamma(b_{\bar{q}}+1)}x^{a_{\bar{q}}}(1-x)^{b_{\bar{q}}}\bar{q}(x),~~~{q=u~\mathrm{or}~d},
\label{eq:02}
\end{aligned}
\end{eqnarray}
where $n_{\bar{q}}$, $a_{\bar{q}}$ and $b_{\bar{q}}$ are free parameters. This form is not meant to give a detailed description of the quark distributions but to optimize the previous linear form. The $\Gamma(x)$-functions are added  for satisfying the normalization
$$\int_0^1 d x \frac{\Gamma(a_{\bar{q}}+b_{\bar{q}}+2)}{\Gamma(a_{\bar{q}}+1)\Gamma(b_{\bar{q}}+1)}x^{a_{\bar{q}}}(1-x)^{b_{\bar{q}}}=1.$$
In our calculations, we set $a_{\bar{q}}=1.0$ and $b_{\bar{q}}=3.0$ phenomenologically as in Refs.~\cite{Ball:2016spl,Anselmino:2008sga}. We also set $a=10^{-3}$ as a reasonable lower limit for the integrations. Here, we adopt a simple approximation to assume that the scale evolution of modeled polarized PDFs only depends on the unpolarized PDF input, though the evolutions of PDFs are not the same for unpolarized and polarized cases from a strict sense.

 \subsection{Numerical calculations}
 We adopt different forms of sea quark helicity distributions as described by Eqs.~(\ref{eq:01}) and (\ref{eq:02}) in our numerical calculations. The parameters for several different modes of sea and valence quark helicity distributions are given in Table~\ref{table1}. We present our numerical results in Table~\ref{table2} and Table~\ref{table3}.	

 In Table~\ref{table1}, the $N_{\bar{u}/\bar{d}}$ and $n_{\bar{u}/\bar{d}}$ are obtained by fitting experimental data of single-spin asymmetries in W boson production at RHIC~\cite{Adamczyk:2014xyw} and the experimental value of $\Gamma^{p,n}$ in COMPASS~\cite{Adolph:2015saz}. From the table, we know that different modes indicate different cases of $\beta_D$, fitting data and relations that we set in Eqs.~(\ref{eq:01}) and (\ref{eq:02}). $\beta_D$ are the harmonic oscillator scale parameters as we mentioned ahead. For example, $\mathrm{Mode}=1,3,5,7$ are corresponding to the fitting procedures with only the data at RHIC~\cite{Adamczyk:2014xyw}, while $\mathrm{Mode}=2,4,6,8$ are corresponding to the fitting procedures with both the data at RHIC~\cite{Adamczyk:2014xyw} and COMPASS~\cite{Adolph:2015saz}. 	
  		\begin{table}[H]\scriptsize
  			\begin{center}
  				\begin{tabular}{c|c|c|c|c|c|c|c|c|c|c|c}
  					\hline \hline
  					\multirow{3}{*}{Relation} & \multirow{3}{*}{$\mathrm{Mode}$}& \multirow{3}{*}{$\beta_{D}$} & \multirow{3}{*}{Data} & \multicolumn{8}{c}{Parameter} \\
  					\cline{5-12}
  					&& && $N_{\bar{u}}$ &$N_{\bar{d}}$ & $n_{\bar{u}}$ & $n_{\bar{d}}$& $a_{\bar{u}}$& $a_{\bar{d}}$& $b_{\bar{u}}$& $b_{\bar{d}}$\\
  					\hline
  					\multirow{4}{*}{Linear}
  				  	&1&\multirow{2}{*}{330}&$\mathrm{{W^{\pm}}}$& 0.242 & -0.309 & - & - & - & -&-&- \\
  				 &2&& $\mathrm{{W^{\pm}+\Gamma^{p,n}_{1}}}$ & 0.001 & -0.040 & - & - & - & -&-&-\\
  				
  					&3&\multirow{2}{*}{600}& $\mathrm{{W^{\pm}}}$ & 0.254 & -0.440 & - & - & - & -&-&- \\
  				 &4&& $\mathrm{{W^{\pm}+\Gamma^{p,n}_{1}}}$& 0.009 & -0.057 & - & - & - & -&-&- \\
  					\hline
  					\multirow{4}{*}{NLinear} &5&\multirow{2}{*}{330}&$\mathrm{{W^{\pm}}}$& - & - &0.150 & -0.225 & 1.0 & 1.0&3.0&3.0\\
  					&6&& $\mathrm{{W^{\pm}+\Gamma^{p,n}_{1}}}$ & - & - &0.010 & -0.197 & 1.0 & 1.0&3.0&3.0\\ 				
  					&7&\multirow{2}{*}{600}& $\mathrm{{W^{\pm}}}$ & - & - &0.159 & -0.319 & 1.0 & 1.0&3.0&3.0\\
  					&8&& $\mathrm{{W^{\pm}+\Gamma^{p,n}_{1}}}$& - & - &0.100 & -0.276 & 1.0 & 1.0&3.0&3.0\\
  					\hline
  					\hline \hline
  				\end{tabular}
  			\end{center}
  		
  			\caption{\label{table1} Parameters of $\Delta \bar q$.}
  		\end{table}
  \begin{sidewaystable}\footnotesize
  		\begin{tabular}{c|c|c|c|c|c|c|c|c|c|c}
  			\hline \hline
  			\multirow{3}{*}{Relation} & \multirow{3}{*}{$\mathrm{Mode}$}& \multirow{3}{*}{$\beta_{D}$} & \multirow{3}{*}{Data} & \multicolumn{7}{c}{Quantity} \\
  			\cline{5-11}
  			&& && $\Gamma^{p}_{1}$ &$\Gamma^{n}_{1}$& $\Delta\Sigma$& $\Delta u^{+}$& $\Delta d^{+}$& $\Delta\bar{u}$& $\Delta\bar{d}$\\
  			\hline
  			\multirow{4}{*}{Linear}
  			&1&\multirow{2}{*}{330}&$\mathrm{{W^{\pm}}}$& 0.289 & -0.210 & 0.282 &1.64&-1.35&0.396&-0.508\\
  			&2&& $\mathrm{{W^{\pm}+\Gamma^{p,n}_{1}}}$ & 0.163 & -0.057 & 0.380 &0.851&-0.471&0.002&-0.066\\
  			
  			&3&\multirow{2}{*}{600}& $\mathrm{{W^{\pm}}}$ & 0.245 & -0.285& -0.144 &1.521&-1.666&0.415&-0.724\\
  			&4&& $\mathrm{{W^{\pm}+\Gamma^{p,n}_{1}}}$& 0.137 & -0.050& 0.316&0.720&-0.404&0.015&-0.093 \\
  			\hline
  			\multirow{4}{*}{NLinear} &5&\multirow{2}{*}{330}&$\mathrm{{W^{\pm}}}$& 0.186 & -0.078  & 0.391 &0.991&-0.599&0.071&-0.130\\
  			&6&& $\mathrm{{W^{\pm}+\Gamma^{p,n}_{1}}}$ & 0.159 & -0.078 & 0.290 &0.857&-0.567&0.005&-0.114\\ 				
  			&7&\multirow{2}{*}{600}& $\mathrm{{W^{\pm}}}$ &0.154&-0.083& 0.256&0.842&-0.585&0.075&-0.184\\
  			&8&& $\mathrm{{W^{\pm}+\Gamma^{p,n}_{1}}}$& 0.145 & -0.075& 0.250 &0.786&-0.536&0.048&-0.159\\ 	
  \hline		
  			\multirow{2}{*}{Parametrization}
  			&-&\multirow{2}{*}&NNPDFpol1.1~\cite{Nocera:2014gqa} &-&-&$0.25\pm0.10$&$0.76\pm0.04$&$-0.41\pm0.04$&$0.04\pm0.05$&$-0.09\pm0.05$\\
             &-&-&DSSV08~\cite{Nocera:2014gqa} &-&-&$+0.366^{+0.042}_{-0.062}\,(+ 0.124)$&$+0.793^{+0.028}_{-0.034}\,(+0.020)$&$-0.416^{+0.035}_{-0.025}\,(- 0.042)$&$+0.028^{+0.059}_{-0.059}\,(+ 0.008)$&$-0.089^{+0.090}_{-0.080}\,(- 0.026)$\\
  			\hline \hline
  		\end{tabular}
  \caption{\label{table2} Quantities from model calculations at $Q=\sqrt{10}$~GeV.}
   		 \end{sidewaystable}

  \begin{table}[H]\scriptsize
  	\begin{center}
  		\begin{tabular}{c|c|c|c|c|c|c|c}
  			\hline \hline
  			\multirow{3}{*}{Relation} & \multirow{3}{*}{$\mathrm{Mode}$}& \multirow{3}{*}{$\beta_{D}$} & \multirow{3}{*}{Data} & \multicolumn{4}{c}{Quantity} \\
  			\cline{5-8}
  			&& && $\Gamma^{p}_{1}$ &$\Gamma^{n}_{1}$& $\Gamma^{p}_{1v}$ &$\Gamma^{n}_{1v}$ \\
  			\hline
  			\multirow{4}{*}{Linear}
  			&1&\multirow{2}{*}{330}&$\mathrm{{W^{\pm}}}$& 0.275 & -0.182 &0.172&-0.027\\
  			&2&& $\mathrm{{W^{\pm}+\Gamma^{p,n}_{1}}}$ & 0.166& -0.052&0.172&-0.027\\
  			
  			&3&\multirow{2}{*}{600}& $\mathrm{{W^{\pm}}}$ & 0.231 & -0.244&0.141&-0.010\\
  			&4&& $\mathrm{{W^{\pm}+\Gamma^{p,n}_{1}}}$& 0.138 & -0.043&0.141&-0.010\\
  			\hline
  			\multirow{4}{*}{NLinear} &5&\multirow{2}{*}{330}&$\mathrm{{W^{\pm}}}$& 0.187 & -0.076&0.172&-0.027  \\
  			&6&& $\mathrm{{W^{\pm}+\Gamma^{p,n}_{1}}}$ & 0.161 & -0.076&0.172&-0.027  \\ 				
  			&7&\multirow{2}{*}{600}& $\mathrm{{W^{\pm}}}$ &0.153&-0.081&0.141&-0.010\\
  			&8&& $\mathrm{{W^{\pm}+\Gamma^{p,n}_{1}}}$& 0.144 & -0.073&0.141&-0.010\\ 			
  			\hline
  			Experiment&-&-&COMPASS~\cite{Adolph:2015saz}&$0.139\pm0.009$&$-0.041\pm0.012$&-&-\\
  			\hline \hline
  		\end{tabular}
  	\end{center}
  	\caption{\label{table3} Quantities from model calculations at $Q=\sqrt{3}$~GeV.}
  \end{table}

Comparing the results between DSSV/NNPDFpol parametrizations~\cite{Nocera:2014gqa,deFlorian:2009vb} and our modes, we can see that different parametrizations can predict different values of $\Delta\bar q$, especially for $\Delta\bar u$. But the basic signs of sea quark distributions are the same. What is more, the sums of polarized distribution functions as shown in our modes are consistent with the parametrizations, while the sea polarized parts seem to be larger. Thus $\Delta\Sigma$, $\Delta q^{+} ~(q=u/d)$ and $\Delta \bar q ~(q=u/d)$ in our study are roughly consistent with those from NNPDFpol1.1~\cite{Nocera:2014gqa} and DSSV08~\cite{deFlorian:2009vb}) parametrizations in Table~\ref{table2}.

In Tables~\ref{table1}-~\ref{table3}, parameters with the linear and nonlinear relations are used to distinguish between different forms of sea quark helicity distributions as in Eqs.~(\ref{eq:01}) and  (\ref{eq:02}). The parameter $\beta_{D}=330$~MeV, which corresponds to $\mathrm{Mode}=1,2,5,6$, and the parameter $\beta_{D}=600$~MeV, which corresponds to $\mathrm{Mode}=3,4,7,8$, are used in the BHL wave function. They have the same magnitude of the quark mass, as a characterization of the transverse momenta of the quark and the diquark. $\beta_{D}=330$~MeV is chosen according to Ref.~\cite{Ma:1996np}, and it has been widely used in describing the baryon properties with only the valence contents considered. We reset a relatively crude $\beta_{D}=600$~MeV by hand as an attempt to reflect possible effects due to the change of the valence helicity distributions in the nucleon by including the sea quark contributions.

As for $W^{+}$ in Eq.~(\ref{eq:1}), the contribution of $\Delta\bar{d}$ is larger than that of $\Delta u$, due to $\bar d\ll u$. Similarly, $\Delta\bar{u}$ plays an important role on $A^{W^{-}}_{L}$ as shown in Eq.~(\ref{eq:2}). In Ref.~\cite{Chen:2005jsa}, the contributions of sea quark helicity distributions are neglected, and such cases correspond to the black dotted curves marked by $\mathrm{Mode=1,2,5,6}$ in our figures. In our work, we also consider the sea quark helicity distributions, just as the solid black curves shown in the figures. Besides, to exam the contributions from up and down sea quark helicity distributions, we calculate $A^{W^{\pm}}_{L}$ by setting one of sea quark helicity distributions $\Delta \bar q=0$, e.g., the black dashed curves represent the contributions from $\Delta \bar d \ne 0$ with $\Delta \bar u=0$; while the black dot-dashed curves stand for the contributions from $\Delta \bar u \ne 0$  with $\Delta \bar d=0$.

\begin{figure}[H]
	\begin{center}
		 \subfigure[$W^{-}$.]{\includegraphics[width=0.45\textwidth]{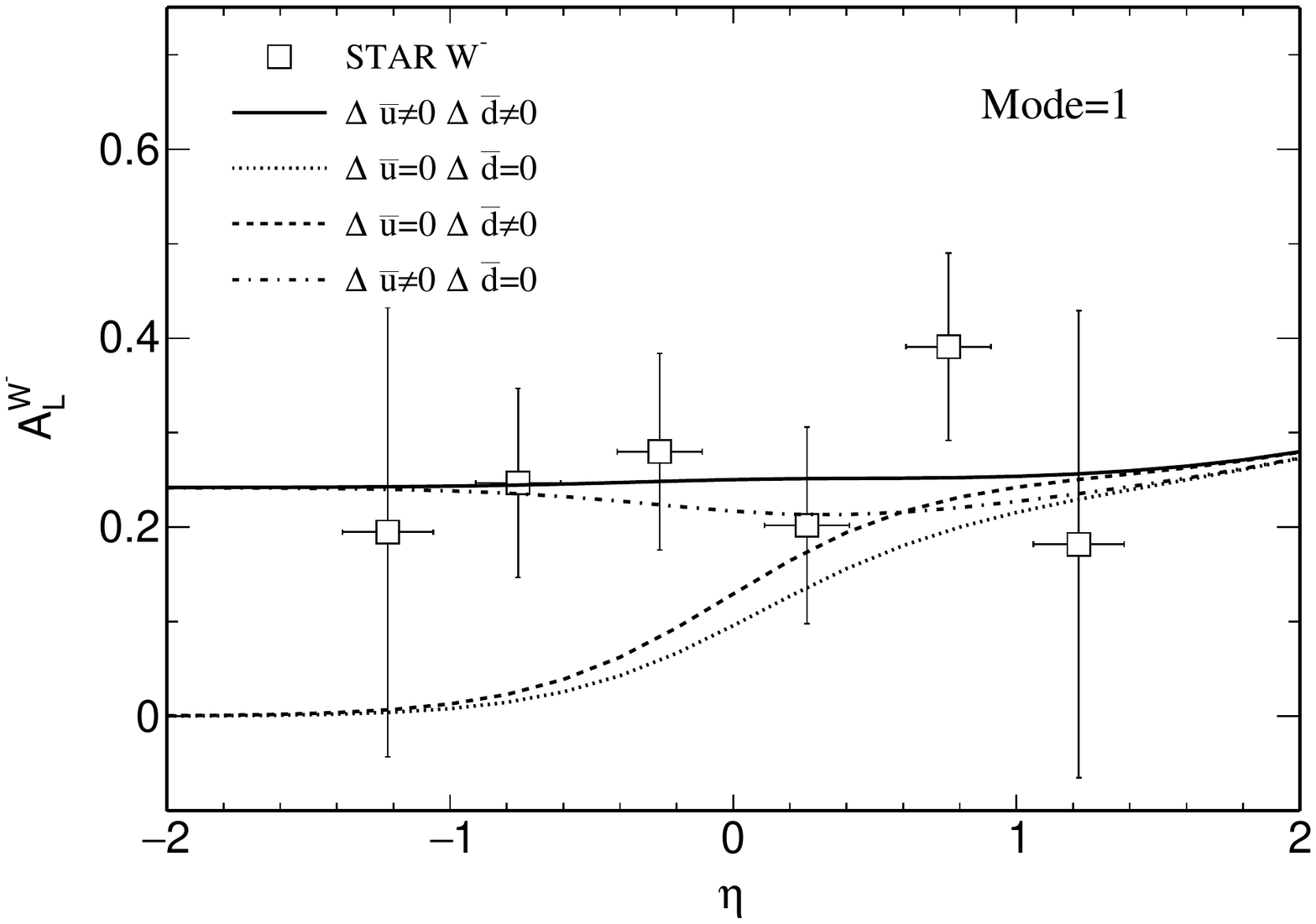}\includegraphics[width=0.45\textwidth]{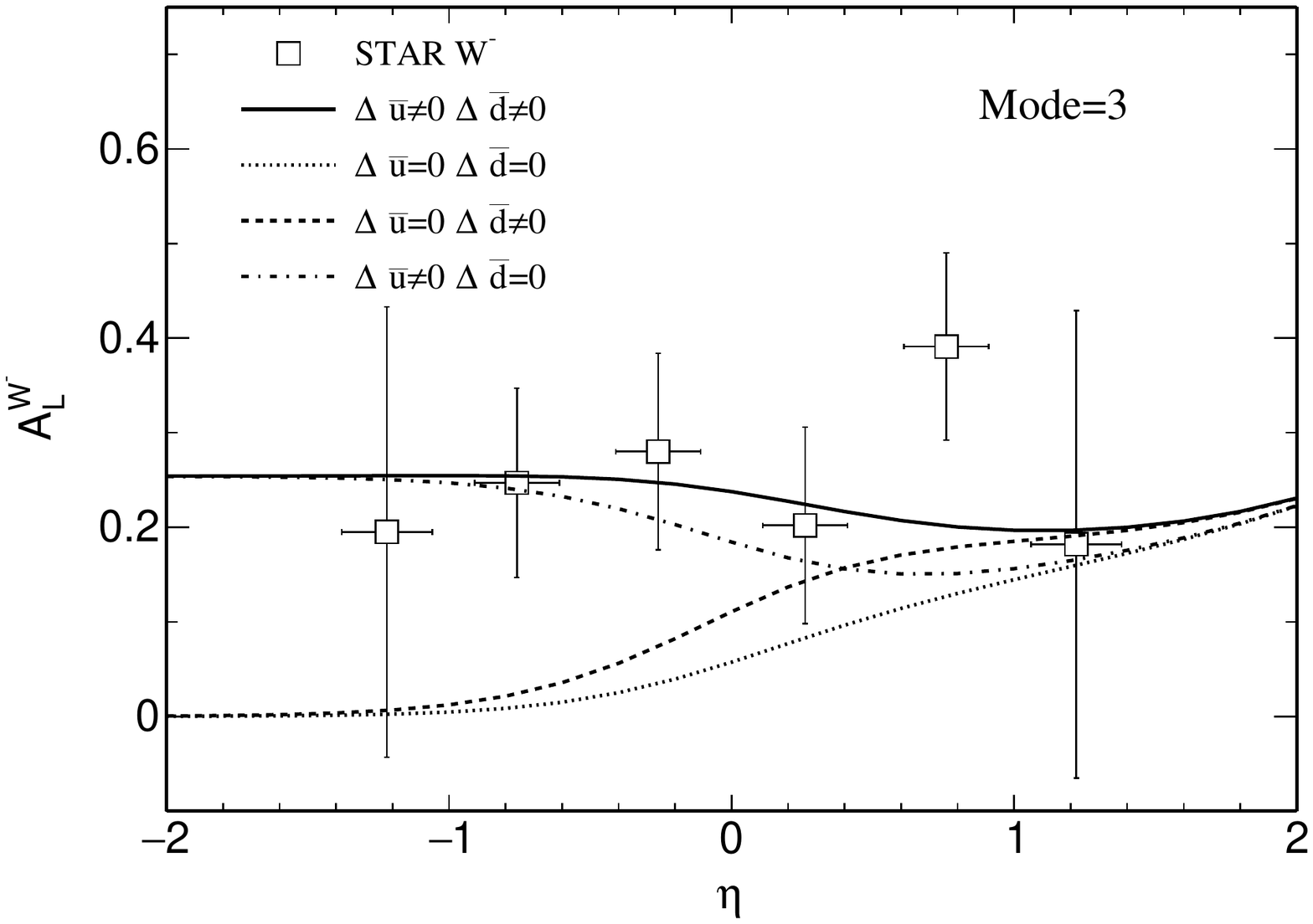}}
		 \subfigure[$W^{+}$.]{\includegraphics[width=0.45\textwidth]{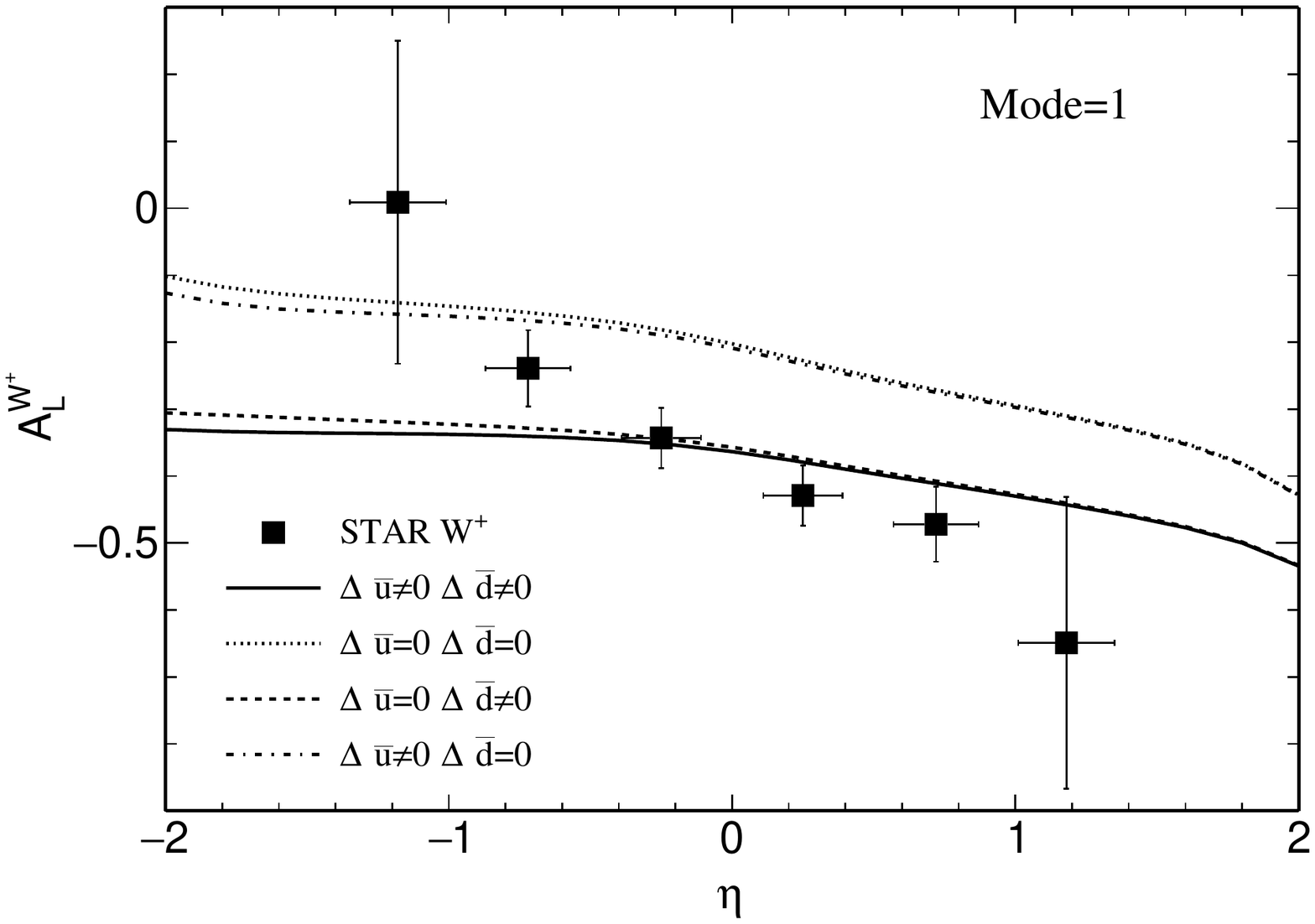}\includegraphics[width=0.45\textwidth]{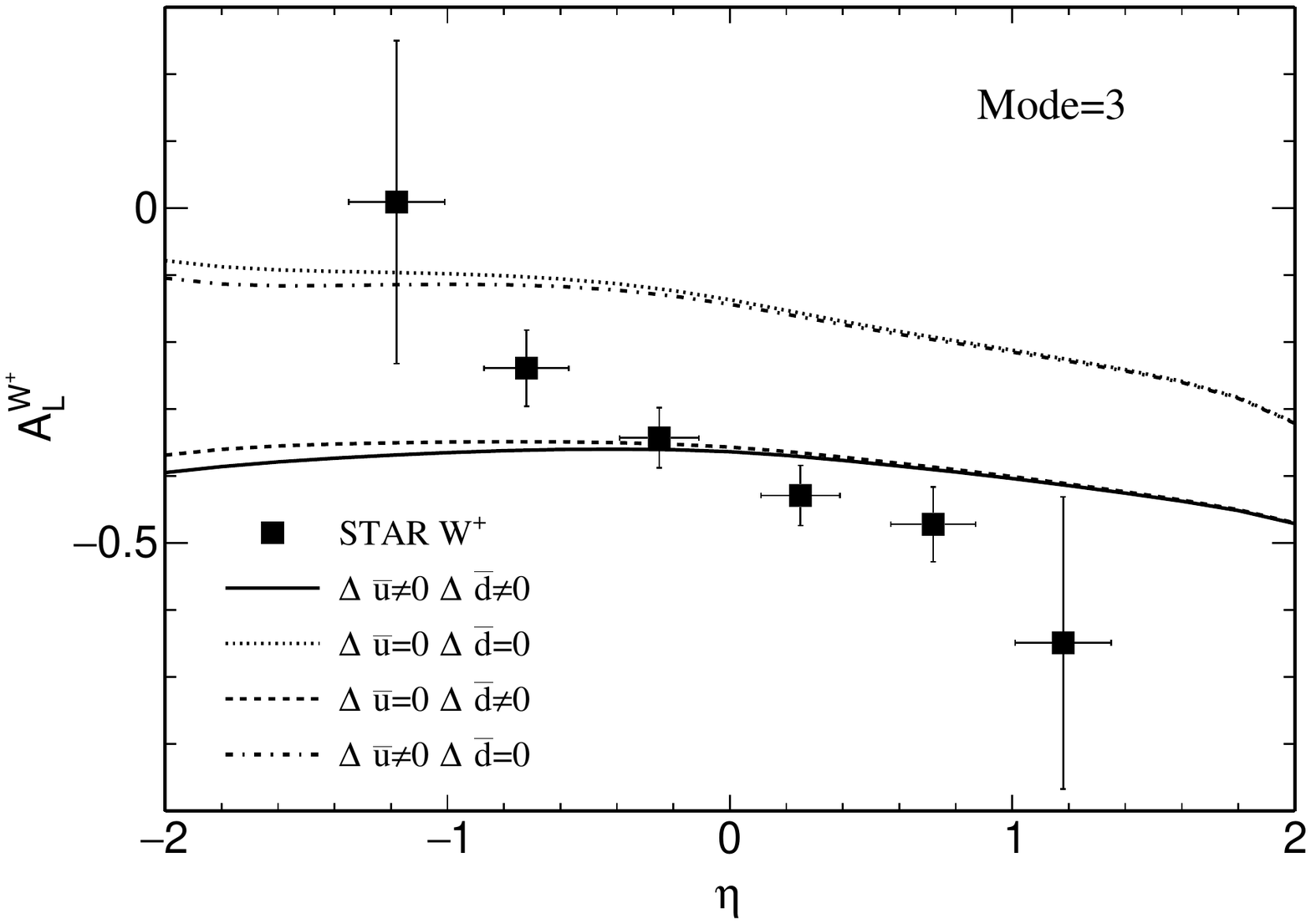}}
	\end{center}
	\vspace{-0.5cm}
	\caption{\label{fig:21} The results of $A^{W^{\pm}}_{L}$ at $Q=M_W/2$~GeV using the linear relation in Eq.~(\ref{eq:01}). $\mathrm{Mode=1}$ and $\mathrm{Mode=3}$ correspond to $\beta_D=330$~MeV and $\beta_D=600$~MeV respectively in the qD model. in the qD model. Both of them are calculated without the constraints of $\Gamma^{p,n}_{1}$. }
\end{figure}

\begin{figure}[H]
	\begin{center}
		 \subfigure[$W^{-}$.]{\includegraphics[width=0.45\textwidth]{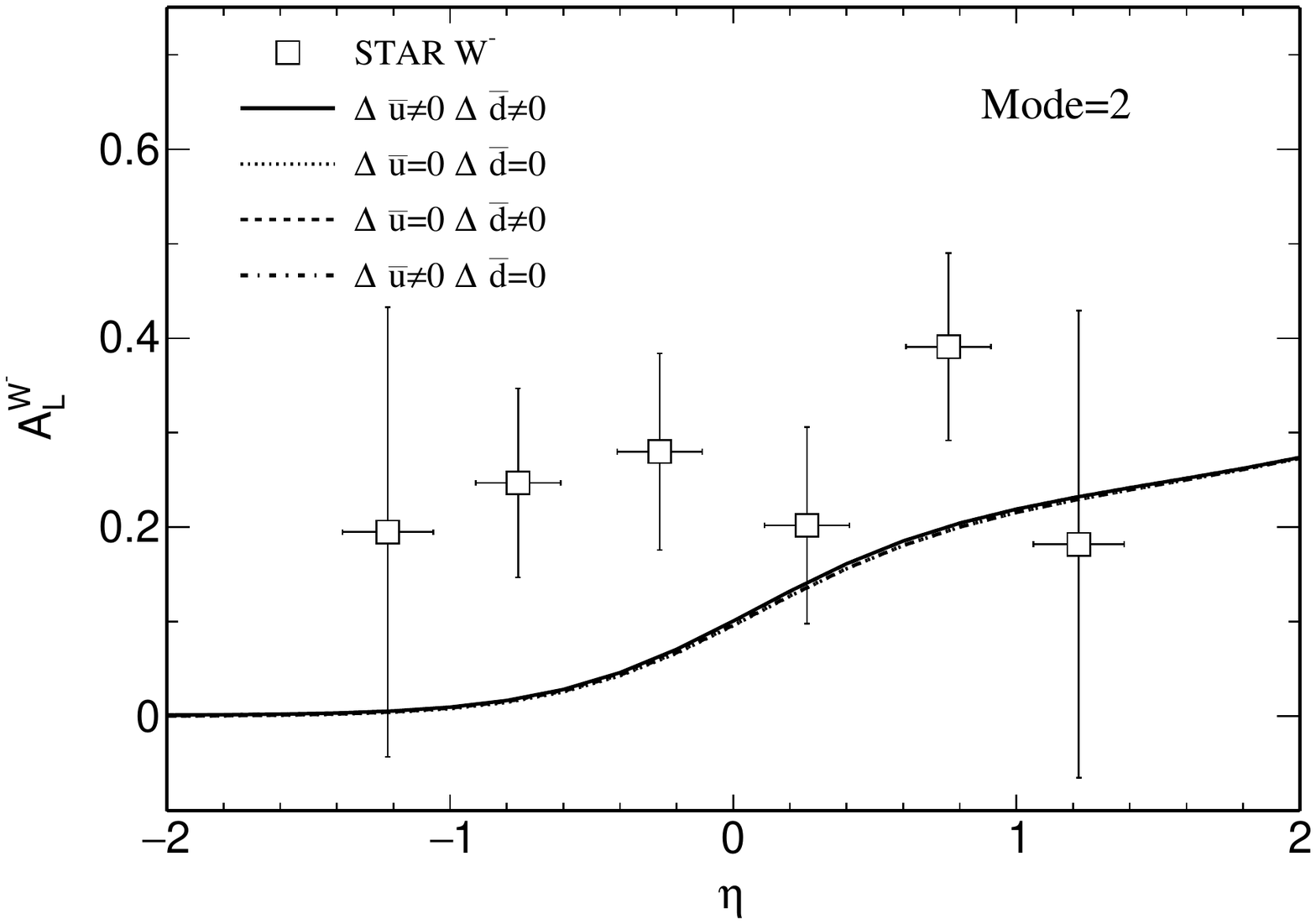}\includegraphics[width=0.45\textwidth]{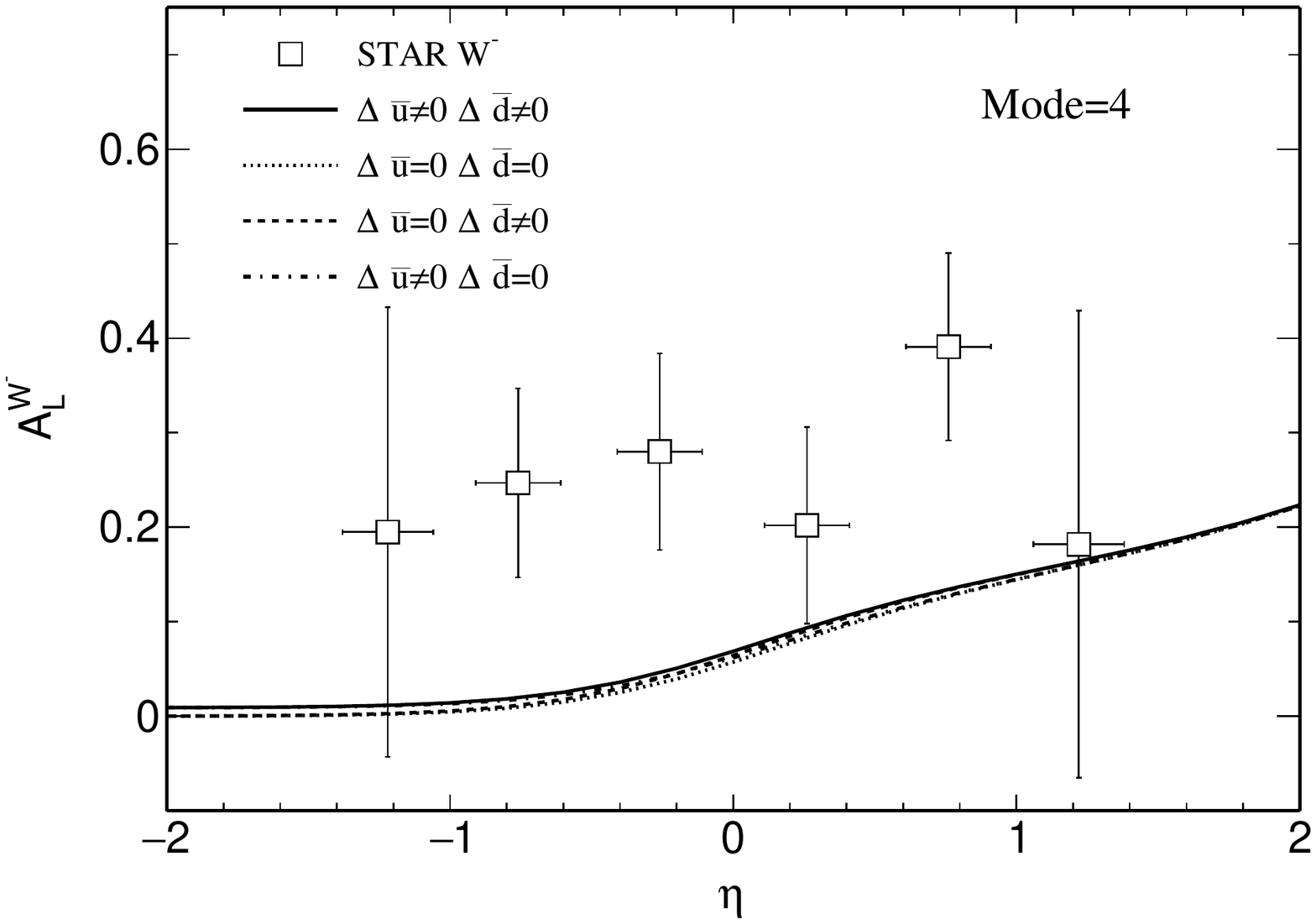}}
		 \subfigure[$W^{+}$.]{\includegraphics[width=0.45\textwidth]{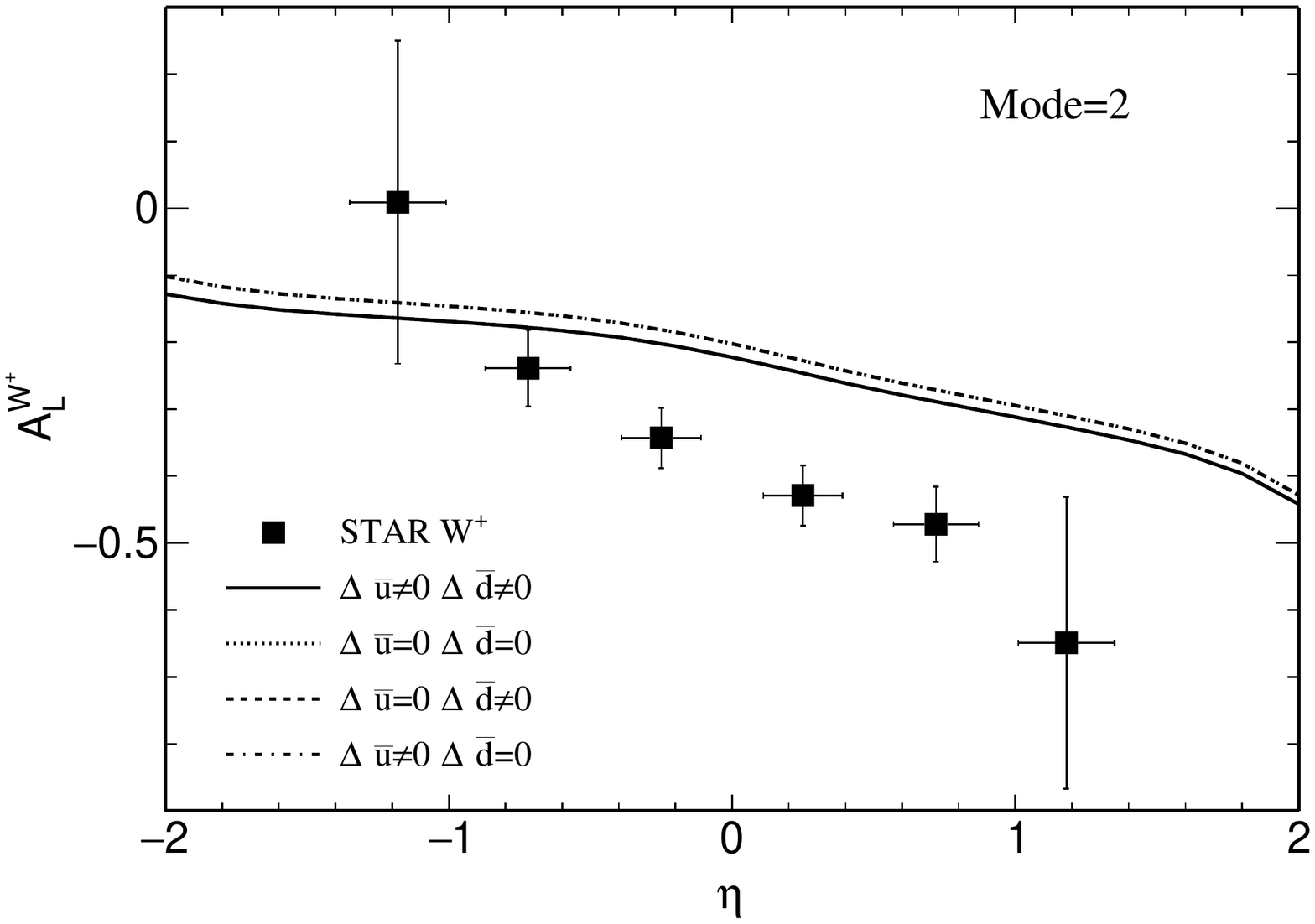}\includegraphics[width=0.45\textwidth]{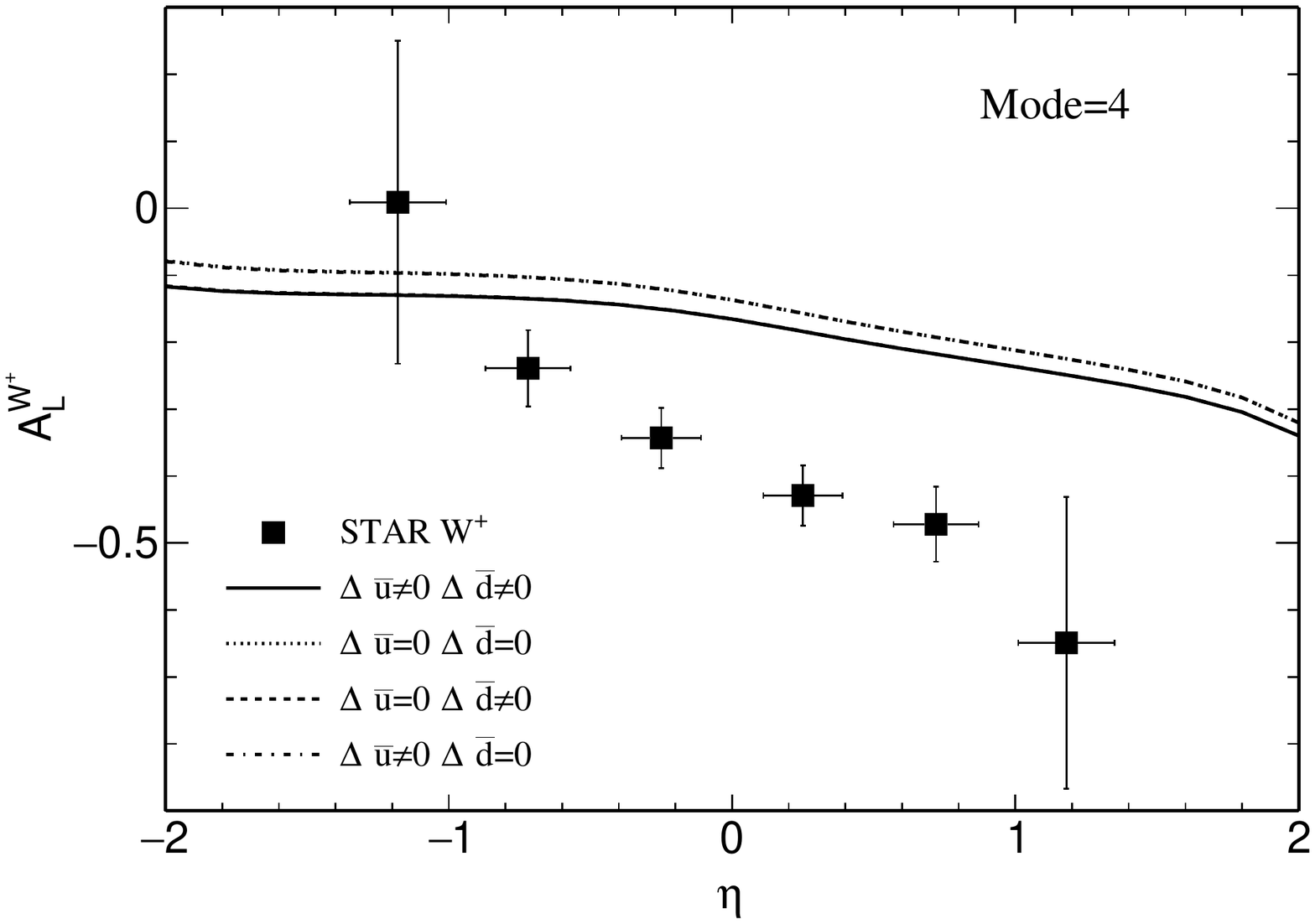}}
	\end{center}
	\vspace{-0.5cm}
	\caption{\label{fig:22} The results of $A^{W^{\pm}}_{L}$ at $Q=M_W/2$~GeV using the linear relation in Eq.~(\ref{eq:01}). $\mathrm{Mode=2}$ and $\mathrm{Mode=4}$ represent that $\beta_D=330$~MeV and $\beta_D=600$~MeV in the qD model. Both of them are calculated with the constraints of $\Gamma^{p,n}_{1}$.}
\end{figure}

\begin{figure}[H]
	\begin{center}
		 \subfigure[$W^{-}$.]{\includegraphics[width=0.45\textwidth]{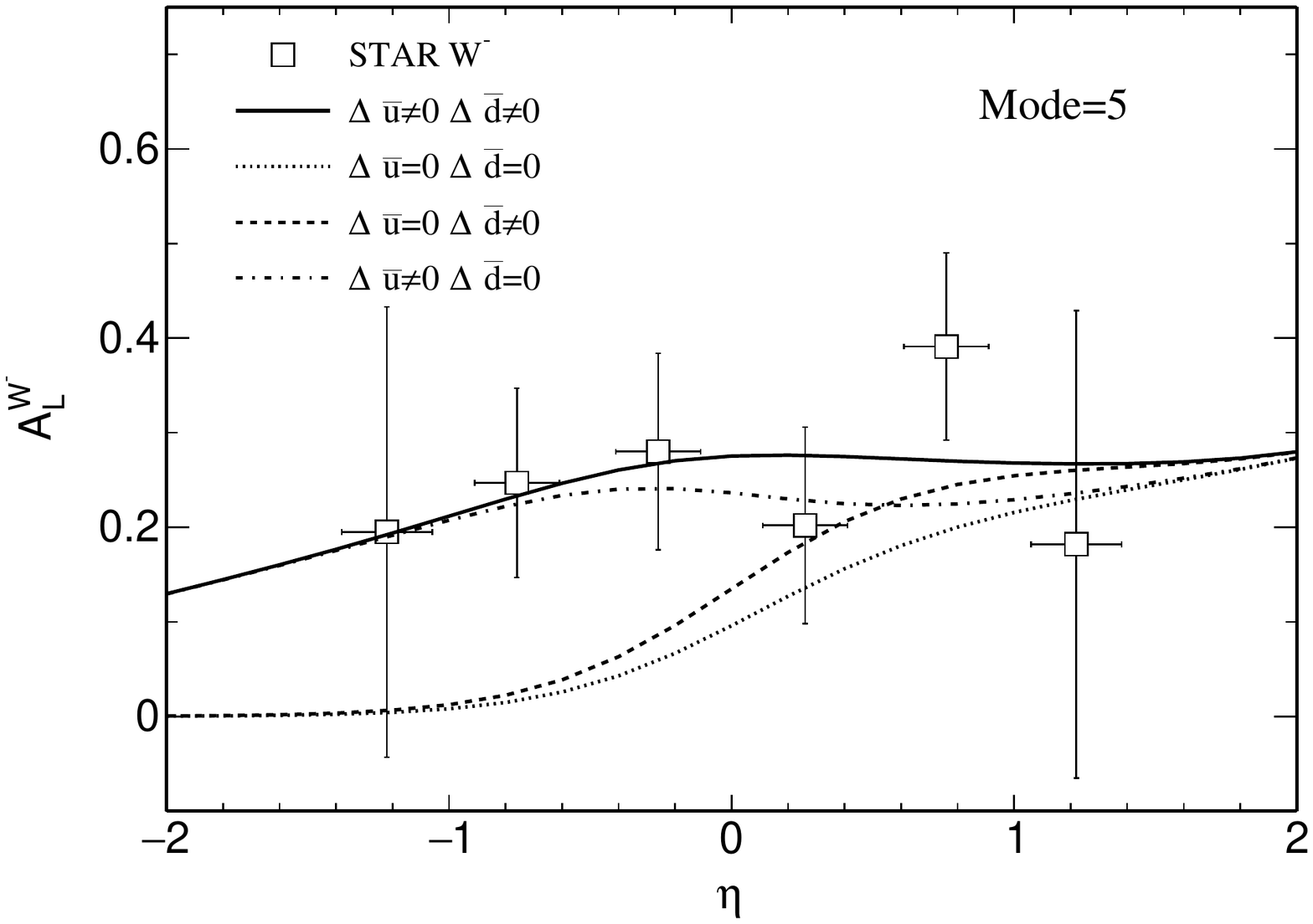}\includegraphics[width=0.45\textwidth]{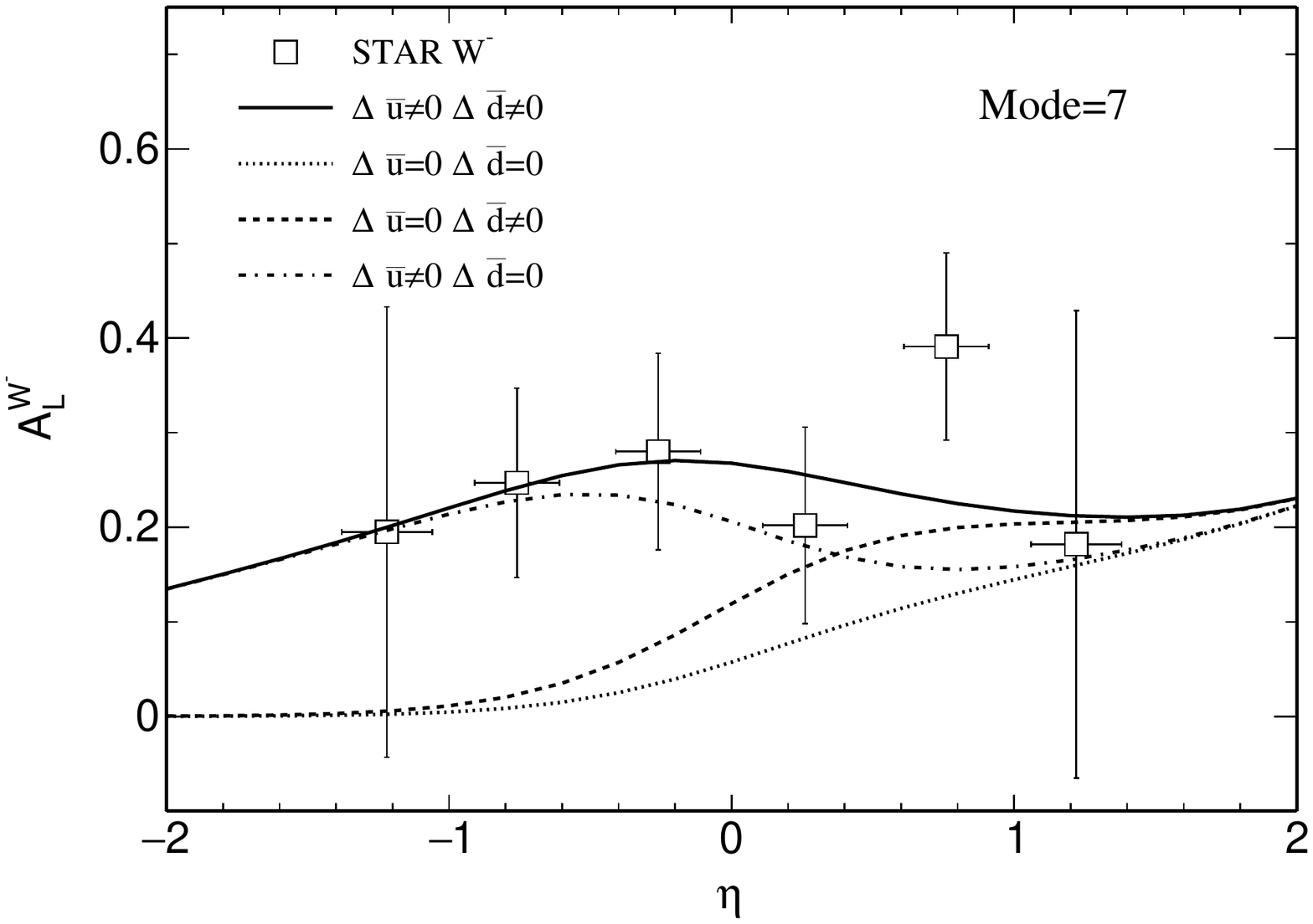}}
		 \subfigure[$W^{+}$.]{\includegraphics[width=0.45\textwidth]{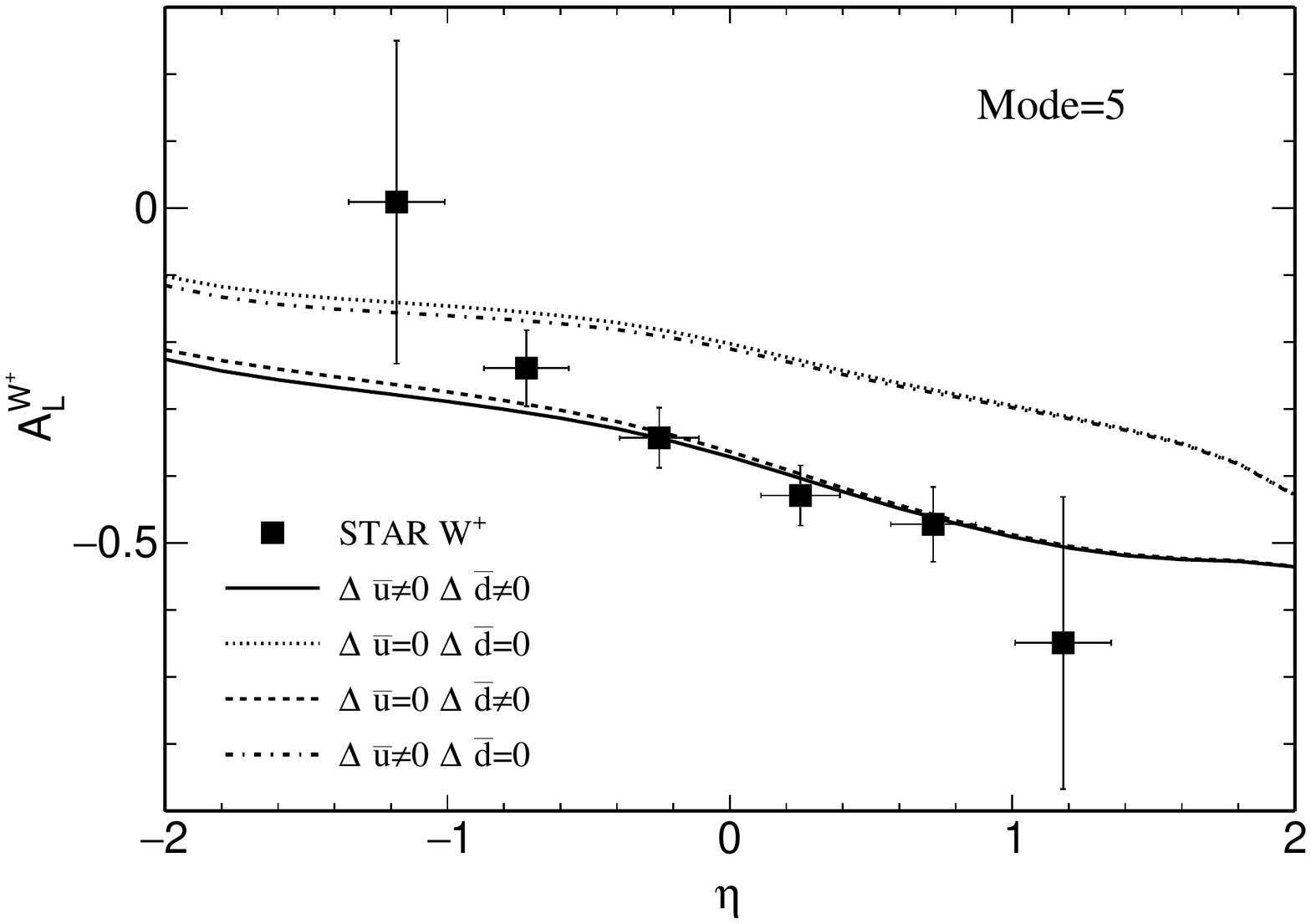}\includegraphics[width=0.45\textwidth]{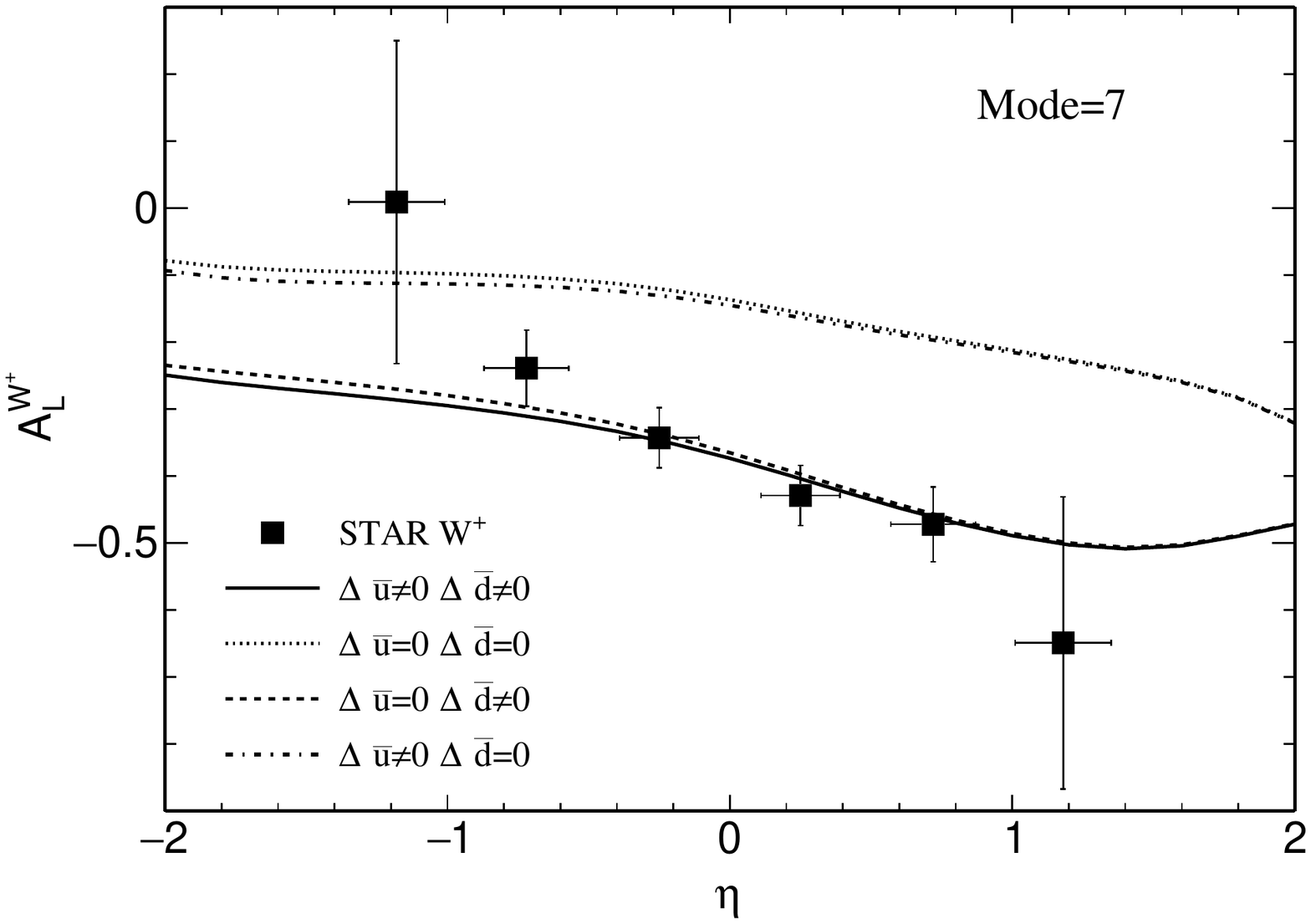}}
	\end{center}
	\vspace{-0.5cm}
	\caption{\label{fig:23} The results of $A^{W^{\pm}}_{L}$ at $Q=M_W/2$~GeV using the nonlinear relation in Eq.~(\ref{eq:02}). $\mathrm{Mode=5}$ and $\mathrm{Mode=7}$ correspond to $\beta_D=330$~MeV and $\beta_D=600$~MeV respectively in the qD model. Both of them are calculated without the constraints of $\Gamma^{p,n}_{1}$.}
\end{figure}

\begin{figure}[H]
	\begin{center}
		 \subfigure[$W^{-}$.]{\includegraphics[width=0.45\textwidth]{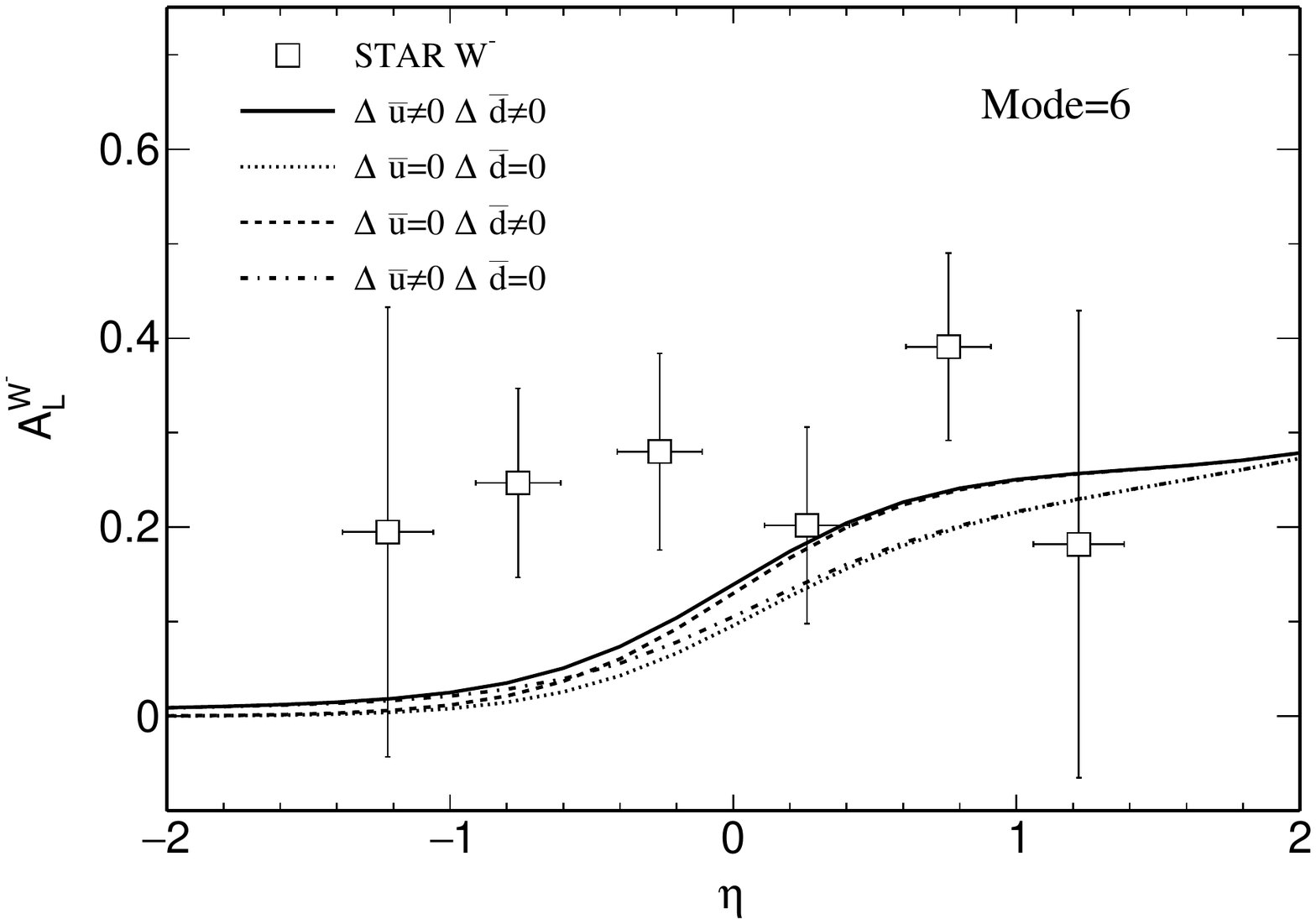}\includegraphics[width=0.45\textwidth]{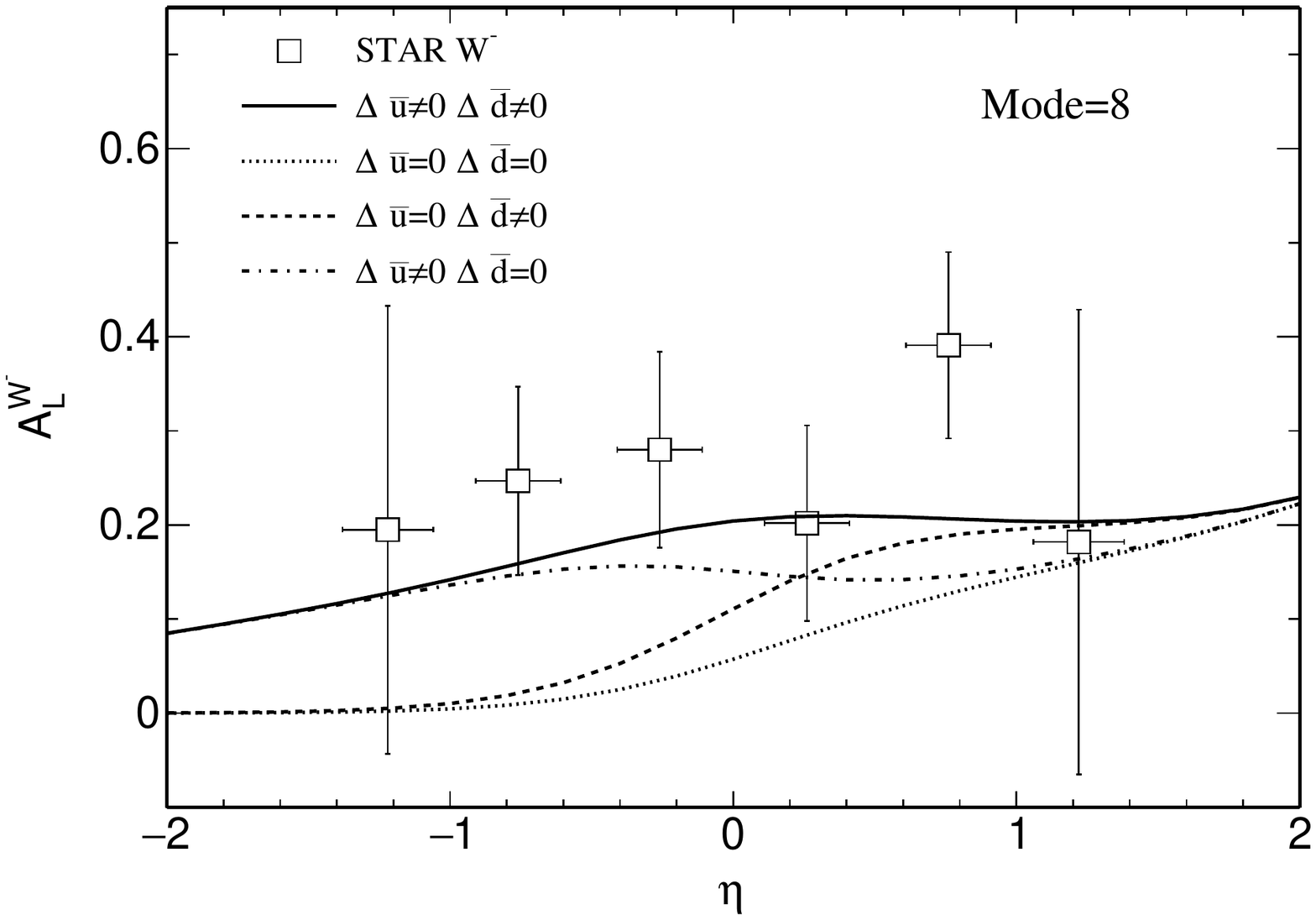}}
		 \subfigure[$W^{+}$.]{\includegraphics[width=0.45\textwidth]{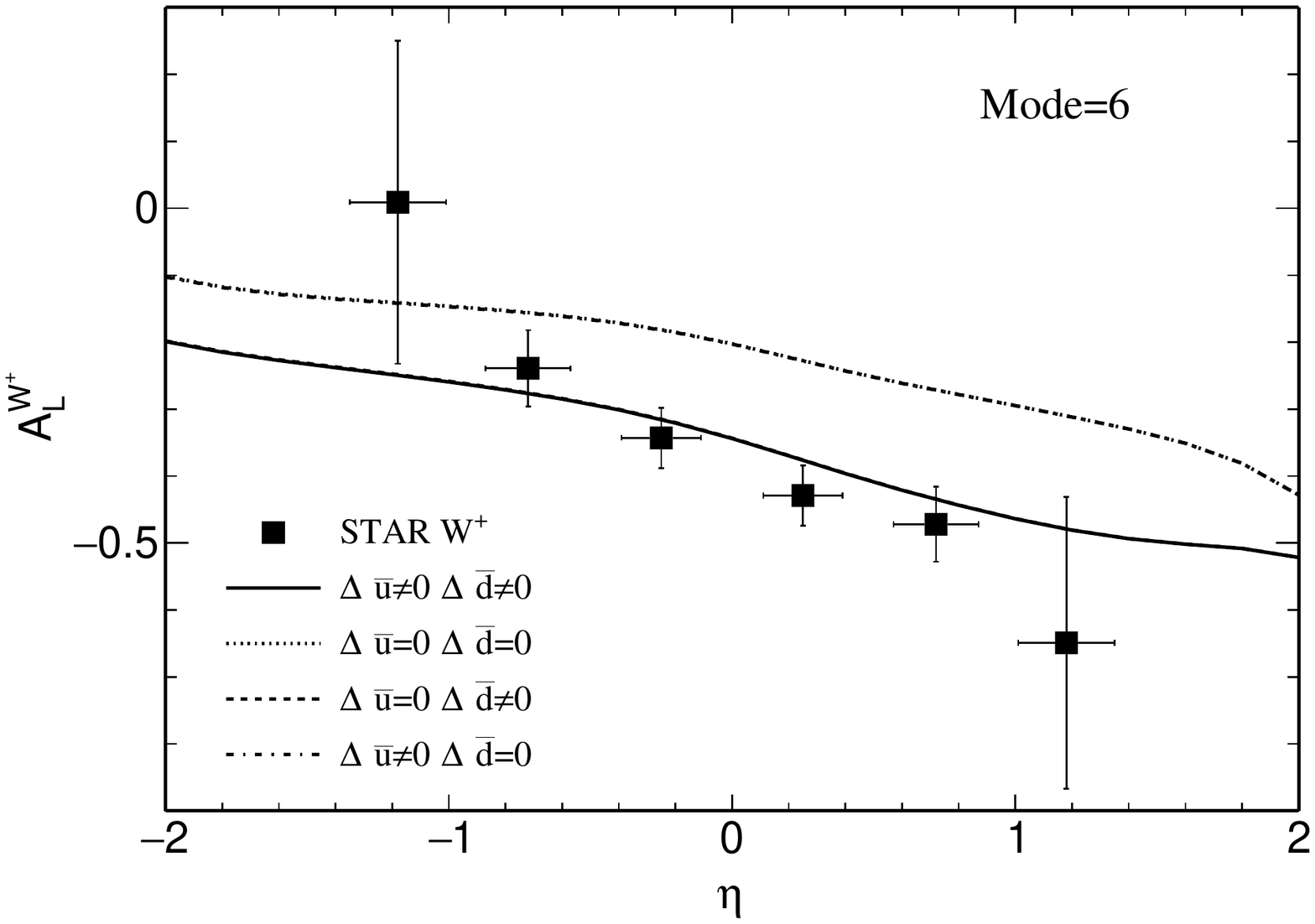}\includegraphics[width=0.45\textwidth]{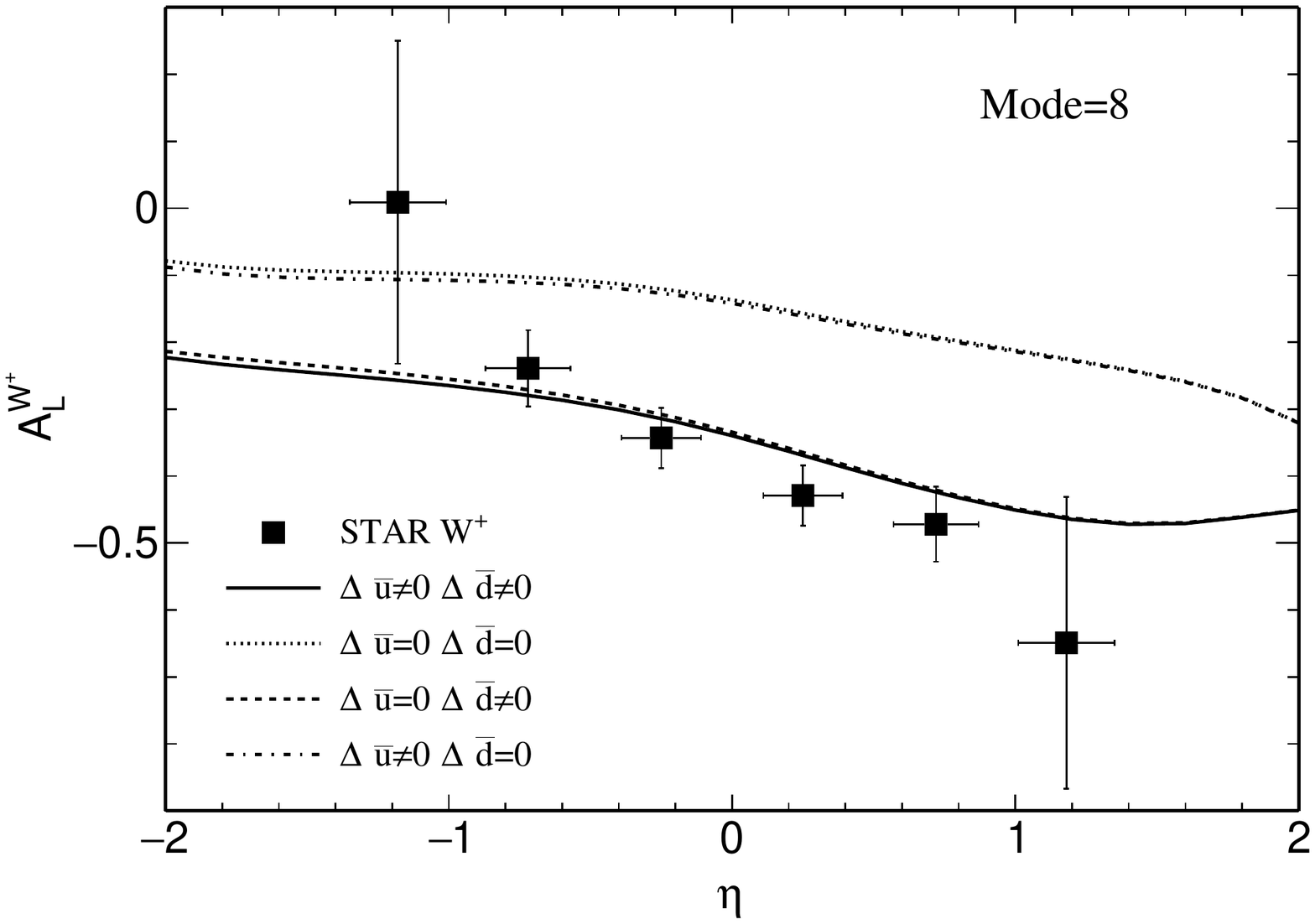}}
	\end{center}
	\vspace{-0.2cm}
	\caption{\label{fig:24} The results of $A^{W^{\pm}}_{L}$ at $Q=M_W/2$~GeV using the nonlinear relation in Eq.~(\ref{eq:02}). $\mathrm{Mode=6}$ and $\mathrm{Mode=8}$ correspond to $\beta_D=330$~MeV and $\beta_D=600$~MeV respectively in the qD model. Both of them are calculated with the constraints of $\Gamma^{p,n}_{1}$.}
\end{figure}

\begin{figure}[H]
 	\begin{center}
 		\subfigure[$\frac{\Delta\bar q(x)}{\bar q(x)}$.]{\includegraphics[width=0.45\textwidth]{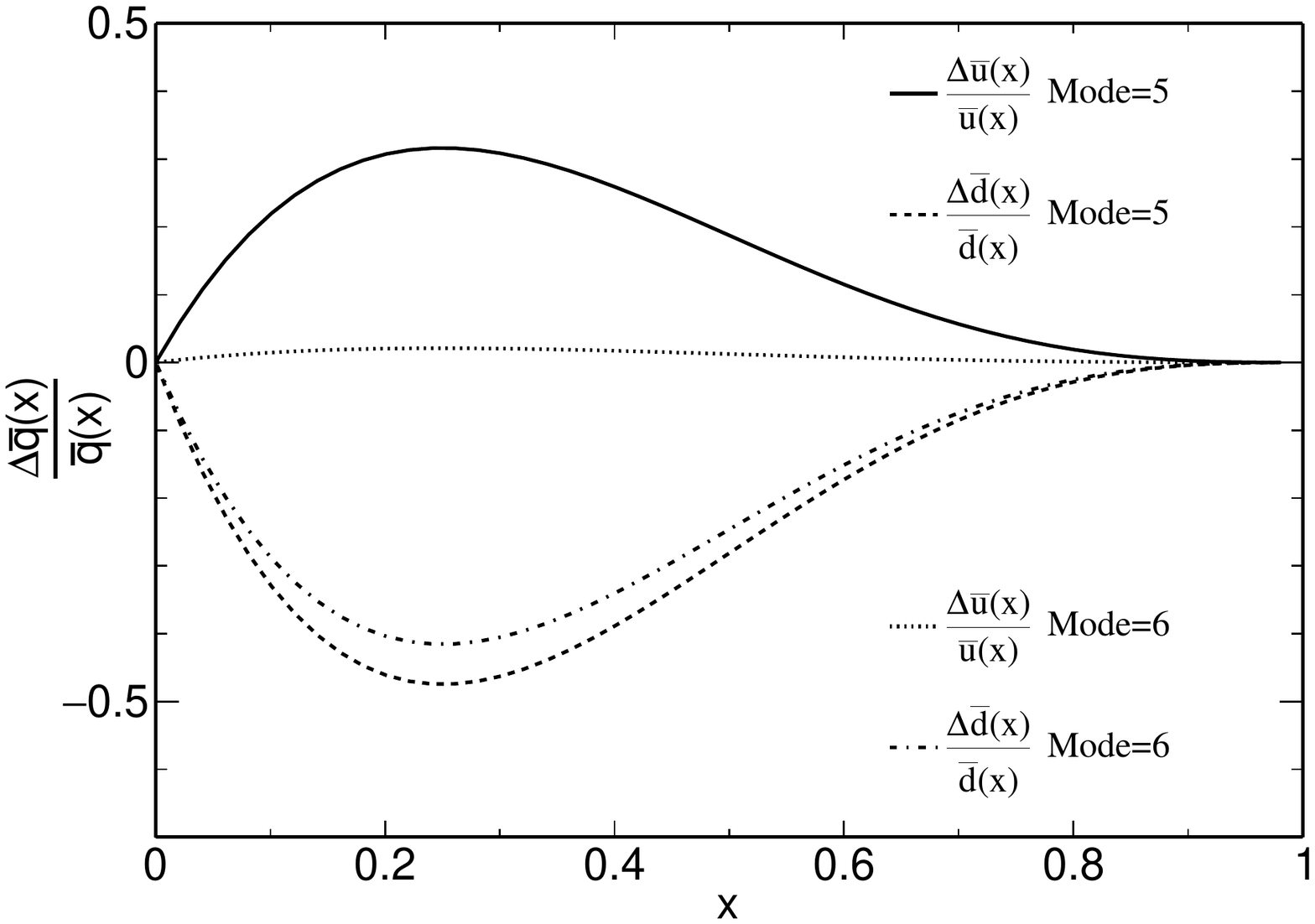}\includegraphics[width=0.45\textwidth]{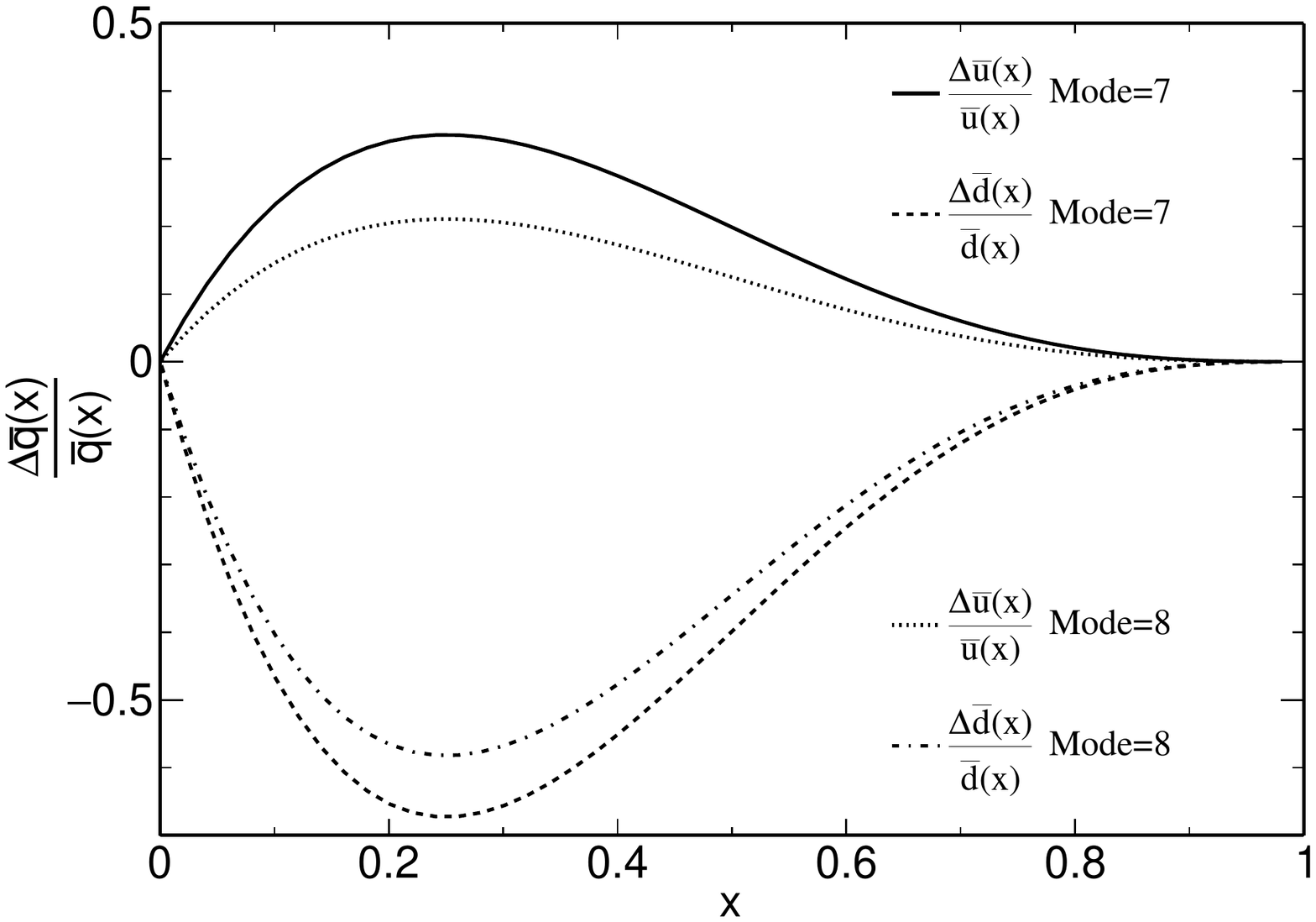}}
 		\subfigure[x$\Delta \bar q(x)$.]{\includegraphics[width=0.45\textwidth]{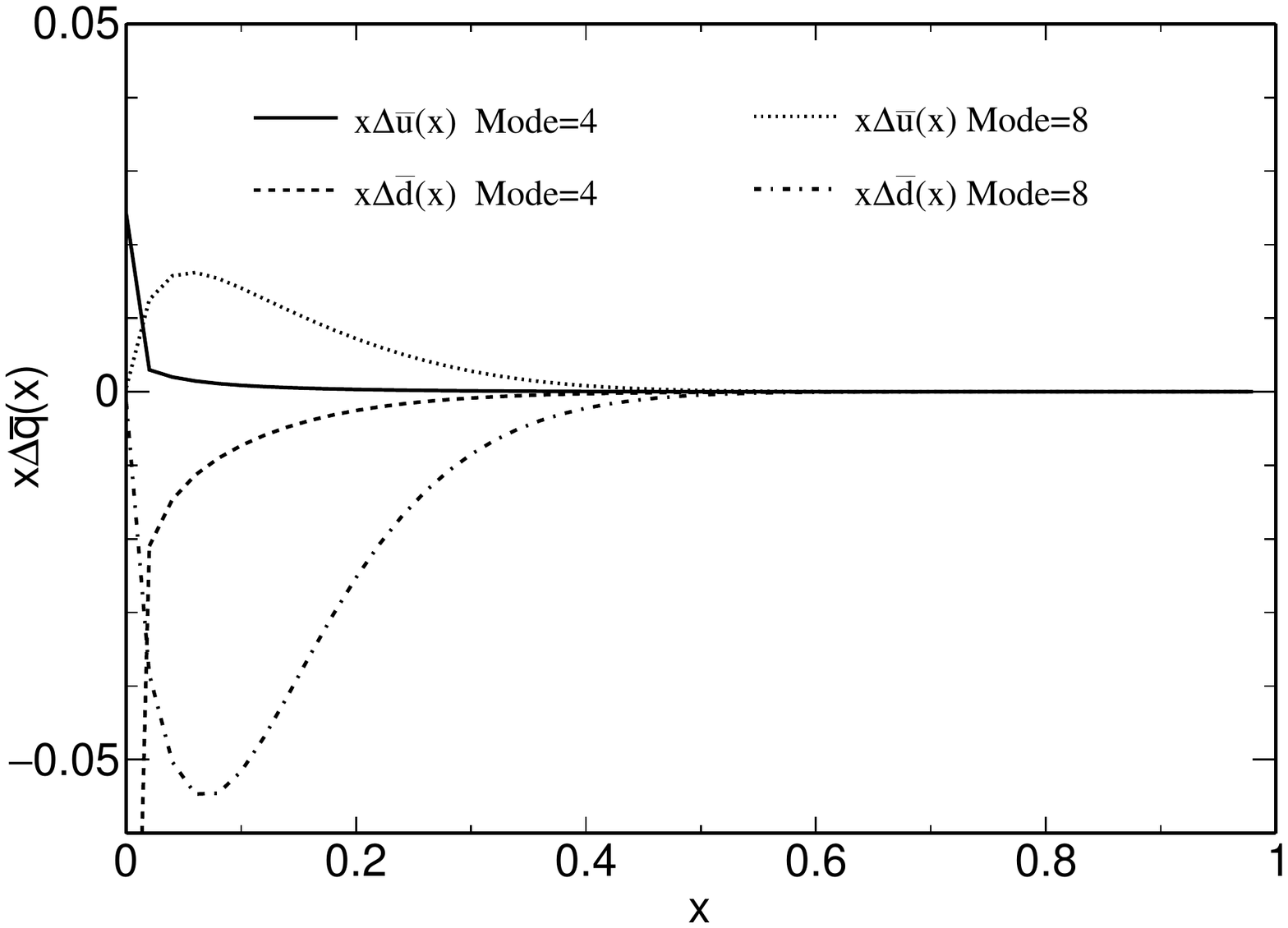}}
 	\end{center}
 	\vspace{-0.5cm}
 	\caption{\label{fig:20} The results of polarized PDFs at $Q=M_W/2$~GeV. }
 \end{figure}

From Table~\ref{table1}, by comparing all of eight different modes we observe that the down sea quarks should be negatively polarized, while the up sea quarks should be positively polarized, for better description of the data. In all modes the ratios of ${\Delta\bar q(x)}/{\bar q(x)}$ with $q=u~\mathrm{or}~d$ satisfy the general relation $|\Delta \bar{q}(x)|\le \bar{q}(x)$ as shown in Fig.~\ref{fig:20}a. In Fig.~{\ref{fig:23}} and Fig.~{\ref{fig:24}}, the calculated $A^{W^{\pm}}_{L}$ can match the data with sizable sea quark helicity distributions, as concluded in~\cite{Ringer:2015oaa}. Besides, a good description of the shape of $A^{W^{-}}_{L}$ at negative $\eta$ depends on positive valued $\Delta \bar{u}$ mainly, while the good reproduction of the shape of $A^{W^{+}}_{L}$ depends on negative valued $\Delta \bar{d}$ mainly. However, in Fig.~{\ref{fig:21}} and Fig.~{\ref{fig:22}}, the results of $A^{W^{\pm}}_{L}$ have a gap with the experimental data due to the unreasonable linear forms of sea quark helicity distributions in Eq.~(\ref{eq:01}).

Besides, with the constraints from $\Gamma^{p,n}_{1}$, which have higher accuracies and reliabilities compared with $A^{W^{\pm}}_{L}$ during our extractions, the values of sea quark helicity distributions become smaller for $\mathrm{Mode=2,4,6,8}$ as shown in Table~\ref{table1}. The reason is that the sums of valence helicity distributions and sea helicity distributions are constrained by $\Gamma^{p,n}_{1}$ while the valence quark helicity distributions already have offered a large value obtained from the qD model, so the sea part turns out to be rather small. Thus $\Gamma^{p,n}_{1}$ and $\Delta\Sigma$ can obtain reasonable values as in Table~\ref{table2} and Table~\ref{table3} from $\mathrm{Mode=2,4,6,8}$. But the small sea quark helicity distributions can not reproduce the shapes of $A^{W^{\pm}}_{L}$ for $\mathrm{Mode=2,4,6}$ as shown in Fig.~{\ref{fig:22}} and Fig.~{\ref{fig:24}}.

From above discussions, we know that $A^{W^{\pm}}_{L}$ need large sizes of sea quark helicity distributions to match the experimental data, while $\Gamma^{p,n}_{1}$ have strong constraints on the sizes of sea quarks helicity distributions.
Due to the inconsistence between $\Gamma^{p,n}_{1}$ and $A^{W^{\pm}}_{L}$ by the sea quark helicity distributions, we try to change both the sea and the valence quark helicity distributions to obtain reasonable results of both $\Gamma^{p,n}_{1}$ and $A^{W^{\pm}}_{L}$ as $\mathrm{Mode=3,4,7,8}$ with adjusted $\beta_{D}=600$~MeV for the valence part of the qD model. We can see that there are little differences for the unpolarized valence quark distributions between different $\beta_{D}$ values because we adopt the parametrization of Eq.~(\ref{eq:q5}). The larger $\beta_D$ has an obvious impact on the valence quark helicity distributions because of the Melosh-Wigner rotation effect. Thus in theoretical calculations, we can reduce the values of valence quark helicity distributions by changing $\beta_{D}=330$~MeV into $\beta_{D}=600$~MeV in the qD model, as shown in Table~\ref{table2} and Table~\ref{table3}. Therefore
these two sets of parameters as in $\mathrm{Mode=4,8}$, with the constraints of $\Gamma^{p,n}_{1}$, can give reasonable results of $\Delta\Sigma$ by adjusting the valence and the sea quark distributions simultaneously. To some degree, the results of $\Gamma^{p,n}_{1}$ and $\Delta\Sigma$ in $\mathrm{Mode=4,8}$ are more reasonable than those in other modes from Table~\ref{table2} and Table~\ref{table3}.
The reason of adjusting $\beta_{D}$ is that the sea quark polarization is usually neglected in previous application of the qD model, so $\beta_{D} \approx330$~MeV
is adopted to reproduce the low energy properties by considering only the valence part of the nucleon~\cite{Ma:1996np,Ma:2002xu,Ma:2002ir,Zhang:2016qqg}. Also the previous extractions of quark helicity distributions are mainly from DIS or SIDIS processes, where the sea quark helicity distributions could not be separated from the total quark helicity distributions accurately. So the valence quark helicity distributions in previous extractions might be overestimated due to the neglect of sea quark polarization.
This implies that we need to reconsider the valence part of the nucleon in the qD model due to the sizable sea quark polarization as suggested by the $A^{W^{\pm}}_{L}$ data.


Even though with the consideration of $\Gamma^{p,n}_{1}$ and $\beta_D$ simultaneously, there still exist large difference for the shapes of $A^{W^{\pm}}_{L}$ between $\mathrm{Mode=4}$ and $\mathrm{Mode=8}$ due to the different forms of sea quark helicity distributions in Eqs.~(\ref{eq:01}) and ~(\ref{eq:02}).
From our calculations, when $\eta_l\in{(-1.2,1.2)}$, the value of $x_1~(x_2)$ satisfies $x_1~(x_2)\in{(0.1,0.5)}$.
Thus to match the shapes of $A^{W^{\pm}}_{L}$ at RHIC, $\Delta\bar q$ must have a large value for especially the parton momentum fraction $x_{n}\in{(0.1,0.5)}$ according to the above discussions. In Fig.~\ref{fig:20}b, we notice that the nonlinear case in Eq.~(\ref{eq:02}) has larger sea quark helicity distributions than the linear form in Eq.~(\ref{eq:01}) when $x_{n}\in{(0.1,0.5)}$. Therefore the linear form of $\Delta\bar q$ should enlarge some orders  in magnitude to match with $A^{W^{\pm}}_{L}$. But the constraints of $\Gamma^{p,n}_{1}$ can never allow $\Delta \bar q$ to enlarge several orders in magnitude. So our results in $\mathrm{Mode=8}$ can give better descriptions than those in $\mathrm{Mode=4}$ for the shapes of $A^{W^{\pm}}_{L}$ .
Thus the $A^{W^{\pm}}_{L}$ data have strong constraints on the explicit forms of sea quark helicity distributions.
Our results of positively polarized $\Delta \bar{u}$ and negatively polarized $\Delta \bar{d}$ are also compatible with a statistical model calculation of parton distributions in~\cite{Bourrely:2013qfa}, where a good description of $W^{\pm}$ asymmetry can be reasonably reproduced.

From the above discussions, we know that the theoretical calculations of $A^{W^{\pm}}_{L}$ could match the experimental data with sizable sea quark helicity distributions. Besides, the $x$-dependent relation could describe the shapes better due to the extra $x$-dependent factors. Additionally, by studying $A^{W^{\pm}}_{L}$ in Drell-Yan process, we note that both of valence and sea quark helicity distributions need to be reconsidered to obtain reasonable descriptions of experimental data.


\section{Summary}\label{sec:ww3}
In summary, we investigate the contributions from the sea quark helicity distributions to the single-spin asymmetries $A^{W^{\pm}}_{L}$ of $W^{\pm}$ bosons in polarized pp collisions. To confront with the experimental data at RHIC, we adopt eight different modes of helicity distributions in our calculations. It is shown that $A^{W^{\pm}}_{L}$ are sensitive to the helicity distributions of quarks, especially the sea quarks. However, the sizes of sea and valence quark helicity distributions are strongly constrained by the experimental data of polarized structure functions and the sum of quark helicities. This study provides an intuitive picture about the role played by the single-spin asymmetries $A^{W^{\pm}}_{L}$ on our understanding of the nucleon spin structure. Therefore further theoretical and experimental studies are needed to explore the helicity distributions of both sea quarks and valence quarks of the nucleon in more details.

\section*{Acknowledgments}
This work is partially supported by National Natural Science Foundation of China (Grant No.~11475006).

\end{document}